\documentclass[12pt]{article}
\pdfoutput=1

\usepackage{cite}
\usepackage{booktabs}
\usepackage[english]{babel}
\usepackage{amsmath,amssymb,amsbsy,amstext, amsthm, simplewick}
\usepackage{hyperref}
\usepackage{graphicx}
\usepackage{amsfonts}
\usepackage{amssymb}
\usepackage[small]{caption}
\usepackage{upgreek}
\usepackage[svgnames,dvipsnames,x11names,table]{xcolor}
\usepackage{multirow} 
\usepackage{geometry}
\usepackage[hang,flushmargin]{footmisc}
\usepackage{bm}
\usepackage{braket}
\usepackage{subcaption}
\usepackage{mathtools}
\usepackage{setspace}
\usepackage{cleveref}
\usepackage{comment}
\usepackage{scalerel}
\usepackage[normalem]{ulem}
\usepackage{slashed}
\usepackage{enumitem}
\usepackage{dsfont}
\usepackage{tikz}
\usetikzlibrary{decorations.markings}
\usetikzlibrary{shapes.misc}
\usetikzlibrary{arrows}
\usetikzlibrary{spy}
\usetikzlibrary{calc}
\usepackage{pgfplots}

\makeatletter
\g@addto@macro\bfseries{\boldmath}
\makeatother

\hypersetup{
    colorlinks=true,
    linkcolor={red!50!black},
    citecolor={blue!50!black},
    urlcolor={blue!80!black}
}

\usepackage{colortbl}

\makeatletter
\newlength{\apb@width}
\newcommand{\autoparbox}[2][c]{\settowidth{\apb@width}{#2}\parbox[#1]{\apb@width}{#2}}

\makeatother

\definecolor{lightgray}{gray}{0.9}

\usepackage[framemethod=default]{mdframed}
\newmdenv[skipabove=7pt,
skipbelow=7pt,
rightline=false,
leftline=false,
topline=false,
bottomline=false,
backgroundcolor=gray!10,
linecolor=gray,
innerleftmargin=5pt,
innerrightmargin=5pt,
innertopmargin=5pt,
innerbottommargin=5pt,
leftmargin=0cm,
rightmargin=0cm,
linewidth=4pt]{eBox}

\usepackage[most]{tcolorbox}
\tcbset{colback=white, colframe=black,
        highlight math style= {enhanced, 
            colframe=red,colback=red!10!white,boxsep=0pt}
        }
\definecolor{light-gray}{gray}{0.95}

\crefname{table}{Table}{Tables}
\crefname{equation}{Eq.}{Eqs.}
\crefname{appendix}{App.}{Apps.}
\crefname{section}{Sec.}{Secs.}
\crefname{figure}{Fig.}{Figs.}

\numberwithin{equation}{section}

\def\beq{\begin{equation}}
\def\eeq{\end{equation}}

\def\bea{\begin{eqnarray}}
\def\eea{\end{eqnarray}}

\def\vp{\varphi_{+}}
\def\vm{\varphi_{-}}
\def\vpm{\varphi_{\pm}}

\def\dvpm{{\dot \varphi}_{\pm}}

\def\d{{\rm d}}

\def\beq{\begin{equation}}
\def\eeq{\end{equation}}
\def\bea{\begin{eqnarray}}
\def\eea{\end{eqnarray}}

\def\d{{\rm d}}

\def\O{{\cal O}}

\def\d{{\rm d}}

\def\H{{\cal H}}

\def\k{{\vec{\scaleto{k}{7pt}}}}

\def\kp{{\!\!\vec{{\scaleto{\,\, k}{7pt}}^{\s\prime}}}}

\def\p{{\vec p}}

\def\x{{\vec x}}

\def\t{\texttt{t}}

\DeclareRobustCommand{\SkipTocEntry}[4]{}

\newcommand{\s}{\hspace{0.8pt}}

\def\dim{\Delta}
\def\dimB{\bar{\Delta}}
\def\dimPhi{\Delta_\phi}
\def\dimBPhi{\bar{\Delta}_\phi}
\def\dimSig{\Delta_\sigma}
\def\dimBSig{\bar{\Delta}_\sigma}

\definecolor{blue3}{RGB}{31, 119, 180}
\definecolor{red3}{RGB}{214, 39, 40}
\definecolor{orange3}{RGB}{255, 127, 14}
\definecolor{green3}{RGB}{44, 160, 44}

\pgfplotsset{compat=1.18}

\begin{document}

\begin{titlepage}
\setcounter{page}{1} \baselineskip=15.5pt 
\thispagestyle{empty}
$\quad$
{\raggedleft CERN-TH-2024-103\par}
\vskip 60 pt

\begin{center}

{\fontsize{20.74}{18} \bf 
Operator Origin of Anomalous \\[7pt] 
Dimensions in de Sitter Space}
\end{center}

\vskip 20pt
\begin{center}
\noindent
{\fontsize{12}{18}\selectfont  Timothy Cohen$^{\s 1,2,3}$, Daniel Green$^{\s 4}$, and Yiwen Huang$^{\s 4}$}
\end{center}

\begin{center}
\vskip 4pt
\textit{ $^1${\small Theoretical Physics Department, CERN, 1211 Geneva, Switzerland}
}
\vskip 4pt
\textit{ $^2${\small Theoretical Particle Physics Laboratory, EPFL, 1015 Lausanne, Switzerland}
}
\vskip 4pt
\textit{ $^3${\small Institute for Fundamental Science, University of Oregon, Eugene, OR 97403, USA}
}
\vskip 4pt
\textit{ $^4${\small Department of Physics, University of California, San Diego,  La Jolla, CA 92093, USA}
}

\end{center}

\vspace{0.4cm}
 \begin{center}{\bf Abstract}
  \end{center}
\noindent 
The late time limit of the power spectrum for heavy (principal series) fields in de Sitter space yields a series of polynomial terms with complex scaling dimensions.
Such scaling behavior is expected to result from an associated operator with a complex dimension.
In a free theory, these complex dimensions are known to match the constraints imposed by unitarity on the space of states. 
Yet, perturbative corrections to the scaling behavior of operators are naively inconsistent with unitary evolution of the quantum fields in dS.
This paper demonstrates how to compute one-loop corrections to the scaling dimensions that appear in the two point function from the field theory description in terms of local operators.
We first show how to evaluate these anomalous dimensions using Mellin space, which has the feature that it naturally accommodates a scaleless regulator.
We then explore the consequences for the Soft de Sitter Effective Theory (SdSET) description that emerges in the long wavelength limit.
Carefully matching between the UV and SdSET descriptions requires the introduction of novel non-dynamical ``operators" in the effective theory.
This is not only necessary to reproduce results extracted from the K\"all\'en-Lehmann representation (that use the space of unitary states directly), but it is also required by general arguments that invoke positivity.

\noindent

\end{titlepage}

\setcounter{page}{2}

\restoregeometry

\begin{spacing}{1.2}
\newpage
\setcounter{tocdepth}{2}
\tableofcontents
\end{spacing}

\setstretch{1.1}

\section{Introduction}

Understanding the behavior of long wavelength fields in de Sitter (dS) space is a stepping stone towards answering a number of questions of immense importance across physics~\cite{Flauger:2022hie,Green:2022hhj}. The primordial density fluctuations that we observe in maps of the universe, such as the CMB, encode the physics of the inflationary era when the background was approximately dS~\cite{Baumann:2009ds,Achucarro:2022qrl}. Making accurate predictions for observations of our own universe relies on our understanding of quantum field theory in these backgrounds~\cite{Baumann:2018muz,Green:2022bre}. Even the assumption that enters most data analyses, that the observables computed during inflation are unaltered by reheating and the subsequent thermal evolution, relies on our understanding of corrections to the evolution of the scalar metric fluctuations to all orders in perturbation theory~\cite{Salopek:1990jq,Senatore:2012ya,Assassi:2012et}. More generally, the quantum nature of the universe in the presence of a non-zero cosmological constant has implications for both the origin and fate of our universe, and it cannot be fully understood without the complete mastery of quantum field theory on a fixed dS background~\cite{Akhmedov:2019cfd,Green:2022ovz}.

Anomalous dimensions of operators in dS is a topic of particular interest for both observational and theoretical applications. Observationally, the scaling dimensions of operators that couple to the inflaton are directly observable via the squeezed limit of inflationary correlators~\cite{Chen:2009we,Baumann:2011nk,Chen:2012ge,Noumi:2012vr}. They break the single field consistency conditions~\cite{Maldacena:2002vr,Creminelli:2004yq} and, as a result, leave a unique imprint in galaxy surveys~\cite{Baumann:2012bc} that is an important observational target~\cite{Achucarro:2022qrl,Green:2022hhj,Chang:2022lrw} for upcoming surveys like SPHEREx~\cite{SPHEREx:2014bgr} and beyond~\cite{DESI:2022lza,Schlegel:2022vrv}. Anomalous dimensions change the scaling behavior of this signal (see \emph{e.g.}~\cite{Green:2013rd,Chen:2016nrs}) and thus the strategy for searches for primordial non-Gaussianity~\cite{Agarwal:2013qta,Gleyzes:2016tdh,Green:2023uyz}. 

Theoretically, the spectrum of operators is thought to provide a complete definition of the dynamics of dS. The emerging perspective~\cite{Starobinsky:1994bd,Strominger:2001pn,Maldacena:2002vr,Senatore:2009cf,Bros:2010rku,Marolf:2010zp,Maldacena:2011nz,Creminelli:2011mw,Mata:2012bx,Anninos:2014lwa,Arkani-Hamed:2015bza,Arkani-Hamed:2018kmz,Baumann:2019oyu,Sleight:2019hfp,Sleight:2019mgd,Baumann:2020dch,Sleight:2021plv,Baumann:2022jpr,Mirbabayi:2019qtx,Green:2020txs,Cohen:2020php,Hogervorst:2021uvp,DiPietro:2021sjt,Cohen:2021fzf,Chakraborty:2023qbp} is that, much like Anti de Sitter (AdS)~\cite{Heemskerk:2009pn,Fitzpatrick:2011ia,Fitzpatrick:2012yx}, the dynamics in dS are encoded in operator mixing, anomalous dimensions, and operator product expansion (OPE) coefficients. In this regard, physics in dS looks increasingly like the more familiar scale invariant systems. Correlation functions between operators are naturally decomposed into factors that behave as power laws of the physical distances between points, which has the natural interpretation in terms of scaling dimensions. Each term then obeys the kinds of symmetry constraints that are familiar from Conformal Field Theory (CFT)~\cite{Ginsparg:1988ui,Cardy:1996xt,DiFrancesco:1997nk,Simmons-Duffin:2016gjk}, and have played a significant role in the development of the cosmological bootstrap~\cite{Baumann:2022jpr}.

Existing calculations nonetheless reveal peculiarities that hint that the physics in dS is not entirely conventional. Starting from a theory of real scalar fields, the natural decomposition into scaling operators yields operators with complex dimensions~\cite{dsrep0,Bargmann:1946me,dsrep1,dsrep2,Sun:2021thf,Penedones:2023uqc}.  Our focus here will be on  ``principal series'' fields, which are heavy fields in $d$ spacial dimensions whose mass $m$ satisfies $4 m^2 > (d H)^2$, where $H$ is the dS Hubble constant.  In this case, the lowest dimension ``operators" that appear in the long distance expansion of the field have dimensions 
\beq
\Delta = \frac{d}{2} + i\nu  
\qquad \text{and} \qquad
\bar \Delta  = d - \Delta = \Delta^* \ ,
\eeq
where $\nu = \sqrt{\frac{m^2}{H^2}-\frac{d^2}{4}}$. However, in theories with self-interactions and/or interactions with additional fields, the operators can acquire anomalous dimensions\cite{Bros:2010rku,Marolf:2010zp}, that take the form
\beq
\Delta = \frac{d}{2}+ \gamma + i\nu  
\qquad \text{and} \qquad
\bar \Delta  = \frac{d}{2}+ \gamma - i\nu = \Delta^* \ .
\eeq
These explicit calculations have found $\gamma >0$, which is also required to maintain positivity of the late-time Fisher information~\cite{Green:2023ids}, see also~\cite{DiPietro:2021sjt,Mirbabayi:2022gnl}. Yet, even with $\gamma >0$, the two point functions of the scaling operators with these dimensions are not real, and linear combinations thereof are not manifestly positive. This is particularly surprising given that the original field is real.  Something is clearly missing from this naive interpretation.

The fundamental challenge is how to interpret the appearance of complex scaling dimensions. When the underlying fields are real, we would expect the correlators to obey basic positivity constraints. In fact, for the range of masses when the dimensions are real, the so-called ``complementary series''  ($4 m^2 < (d H)^2$), positivity of the two point functions are known to place stringent constraints on the mass ranges of consistent theories in dS~\cite{Higuchi:1986py}. For complex dimensions where the real part of the dimension is $d/2$, as required for the unitary states, then positivity is maintained and the theories appear to be well defined. However, when the real part of the dimension is corrected, it would appear that positivity is lost, as the contributions from all the scaling operators are oscillatory. This puzzle is present for either sign of the anomalous dimension, but for negative anomalous dimensions this would furthermore be inconsistent with positivity of the power spectra for the classical density fluctuations that we observe in the late universe~\cite{Green:2023ids}. 

The physical origin of the anomalous dimension $\gamma$ has been given an interpretation in dS (and AdS~\cite{Fitzpatrick:2011hu}) as the decay width of massive particles~\cite{Bros:2010rku,Marolf:2010zp}. As energy is not conserved in these spacetimes, the decay of a particle is not forbidden by kinematics and thus is expected to arise generically. Moreover, this interpretation would explain why $\gamma \geq 0$, which is also consistent with the optical theorem in the flat space limit~\cite{Fitzpatrick:2011hu,DiPietro:2021sjt,Mirbabayi:2022gnl}. Unfortunately, this picture of particle decay does not resolve the question of how to understand the lack of explicit positivity of the cosmological correlators, and other puzzles these anomalous dimensions present.

Much of the recent progress in understanding the long-distance description of dS has been in terms of operators and symmetries. The most natural explanation for anomalous dimensions in this language would be in terms of mixing of operators.  
The formula for these anomalous dimensions can be decomposed into a sum over composite operators that appear in the K\"all\'en-Lehmann representation of the loop integral~\cite{Hogervorst:2021uvp,DiPietro:2021sjt}. Moreover, the largest contributions to this sum arises from the operators whose dimensions are closest to the dimension of the external fields. At first sight, one might imagine this is due to mixing between the fundamental field and the composite operators. Unfortunately, this interpretation does not match our experience with conventional Renormalization Group (RG), where one can neglect mixing between operators with non-degenerate dimensions. Furthermore, even if two operators with well separated dimensions appear to mix, this effect can be removed by a change of basis of the operators (diagonalizing the dilatation matrix). In contrast, the loop corrections studied in this paper generate true anomalous dimensions with the associated logarithmic terms, and they thus cannot be removed in this sense. Given the success in the operator approach to the case of light fields~\cite{Baumgart:2019clc,Green:2020txs,Cohen:2021fzf}, one would like to understand how these ``anomalous dimensions" arise for heavy fields in a way that can be predicted using dimensional analysis.

\begin{figure}[t!]
    \centering
    \includegraphics[width=0.75\textwidth]{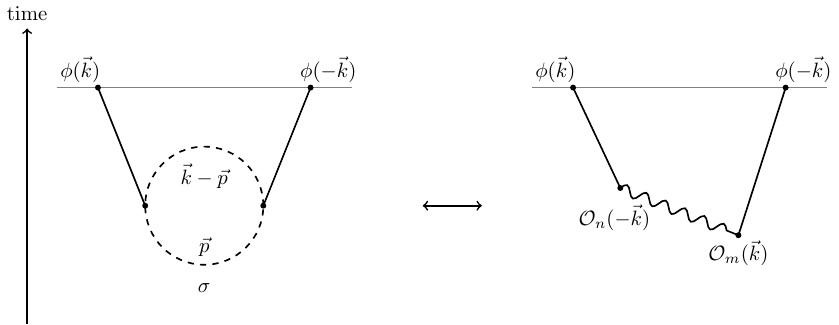} 
    
    \caption{Diagramatic representation of the one-loop power spectrum of a scalar field $\phi$ interacting with another field $\sigma$ in terms of a sum over the exchange of scaling operators, $\langle \O_{n}(\k, \tau) \O_m(\k,\tau') \rangle$.}
    \label{fig:dS_intro}
\end{figure}

In this paper, we will show that these apparent contradictions can be resolved by the appearance of additional non-dynamical scaling operators that appear in the long wavelength description. These operators contribute a unique set of calculable contact terms. These contributions appear in direct calculations of the loop integrals when evaluated in Mellin space~\cite{Premkumar:2021mlz,Qin:2022lva,Qin:2023bjk}. Like cuts in loop diagrams in flat space, these terms arise as imaginary parts of the integral that are not visible from the series expansion. These same contributions are implicit in the K\"all\'en-Lehmann representation as contact terms in the Green's function that multiplies the spectral density. We will show that both methods of calculating produce identical results for the anomalous dimensions.

To expose the nature of these contact terms explicitly, we introduce Soft de Sitter Effective Theory (SdSET), an Effective Field Theory (EFT) description of the long wavelength modes, for principal series fields.  This follows the same approach that was applied for light complementary series fields in~\cite{Cohen:2020php}. In the SdSET description, the matching the short distance and long distance descriptions requires the introduction of non-dynamical operators. The contact terms that are needed to explain the full calculations are not allowed by conformal invariance for any operators with non-zero correlations at separated points. Adding the non-dynamical operators allows us to reintroduce the necessary contact terms while still being consistent with conformal invariance of the full theory. In the process, one finds that the correlation functions of real fields remain positive when the anomalous dimensions are positive. Including additional non-dynamical operators in an effective description has analogies with heavy quark effective theory~\cite{Manohar:1993qn} and soft collinear effective theory~\cite{Liu:2008cc,Bauer:2010cc,Rothstein:2016bsq}, \emph{e.g.}~the need to include Glauber modes.

The additional benefit of the SdSET description is that it permits a description of the anomalous dimensions that is entirely consistent with conventional RG flow. SdSET power counting yields a organization of the interactions as relevant, marginal, or irrelevant; this characterize how corrections grow or decay under time evolution via (dynamical) RG. The new non-dynamical operators introduce relevant deformations of the SdSET action, and give rise to calculable logarithmic corrections within the EFT description.  In this sense, the nontrivial corrections to the long wavelength scaling could have been anticipated entirely from power counting, as one would expect.

The rest of this paper is organized as follows:  In \cref{sec:mellin}, we directly calculate the anomalous dimensions of principal series fields using the Mellin-representation of the mode functions. We confirm our explicit results to the results are equivalent to the calculations the K\"all\'en-Lehmann representation. In \cref{sec:EFT}, we derive SdSET for these fields and explain the origin of anomalous dimensions in the EFT description. We conclude in \cref{sec:con}. In \cref{app:sum}, we provide details how the we sum over the poles in the Mellin representation of the loop integrals.

We will use the following definitions throughout: the length of a vector $\vec{k}$ is denoted $k = |\vec{k}|$. We will define correlation functions without the momentum conserving $\delta$-function as 
\beq
\langle {\cal O}(\k_1) .. {\cal O}(\k_n) \rangle \equiv \langle {\cal O}(\k_1) .. {\cal O}(\k_n) \rangle' (2\pi)^3 \delta\big(\sum \k_i\big) \ .
\eeq
We will be working in a fixed de Sitter background, defined by the metric
\beq
\d s^2 = -\d t^2 + a(t)^2 \d\vec x^{\s 2} = a(\tau)^2 (-\d\tau^2 + \d\vec{x}^{\s 2} ) \ ,
\eeq
where $a(t) = e^{H t}$, $H$ is the constant Hubble rate, and $\tau = -1/(a(\tau)H)$ is the conformal time. We sometimes use the shorthand $[a(t_i) H] = [aH]_i$.  Finally, for SdSET, we will typically work in terms of dimensionless time, $\t \equiv H t$.

\section{Anomalous Dimensions From the UV}\label{sec:mellin}

Loop corrections in de Sitter are notoriously challenging and have been an endless source of confusion~\cite{Akhmedov:2019cfd,Green:2022ovz}. Direct calculations are limited by the need to integrating over Hankel functions, particularly when implementing a regulator like dimensional-regularization. Achieving finite results has often required some type of hard cutoff, which can then introduce unphysical answers~\cite{Senatore:2009cf}.

The introduction of the K\"all\'en-Lehmann representation has enabled a number of calculations of loop corrections, particularly for principal series fields~\cite{Bros:2010rku,Marolf:2010zp,Loparco:2023rug,DiPietro:2023inn,Loparco:2023akg}. However, it can be difficult to interpret the physical origin of some of the corrections that this approach includes automatically. Moreover, because it relies on the representation theory of unitary states in de Sitter, there is no trivial generalization to inflationary cosmologies.

\begin{figure}[t!]
    \centering
    \includegraphics[width=0.3\textwidth]{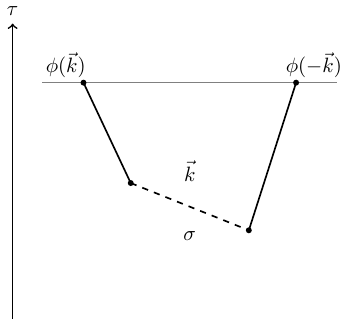} \hskip 20pt
    \includegraphics[width=0.3\textwidth]{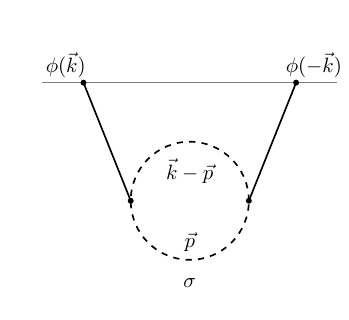}
    
    \caption{The diagram representing the tree ({\it left}) and 1-loop ({\it right}) contributions to the power spectrum of $\phi$ via a quadratic interaction $\H_{\rm int}= m^2 \phi \sigma$ and a cubic interaction $\H_{\rm int} = \lambda H \phi \sigma^2 $ respectively.}
    \label{fig:dS_mix}
\end{figure}

In order to clarify the nature of anomalous dimensions for principal series fields, we will directly calculate two examples cases: mass mixing, and a one-loop correction to the power spectrum, shown in \cref{fig:dS_mix}.
For the one-loop calculation, we will use the Mellin-representation of the Hankel functions to make the calculations tractable and finite without the introduction of hard UV cutoffs~\cite{Premkumar:2021mlz}. The advantage of the Mellin representation is that it provides an efficient packaging of the series representation. We expect to be able to map each contribution expressed in Mellin space onto a corresponding operator statement in the long distance theory. Another benefit of using this description is that $\delta$-function localized terms (in space) are easy to identify and calculate; this will play a major role in deriving the results of this paper.

\subsection{Principal Series in Mellin Space}
Take a scalar field $\phi$ to be in the principal series.  The canonically quantized field operator is
\beq
\hat \phi(\k,\tau)=\bar\phi_k(\tau) \hat a_{\vec{k}}^{\dagger}+\bar\phi^*_k(\tau) \hat a^{\vphantom{\dagger}}_{-\vec{k}} \ ,
\eeq
where $\big[\hat a_{\vec{k}}^{\dagger}, \hat a^{\vphantom{\dagger}}_{\vec{k}^{\prime}}\big]=(2 \pi)^3 \delta(\k-\kp)$, and
\begin{subequations}
\begin{align}
\bar\phi_k(\tau)&=\frac{\sqrt{\pi}}{2} e^{\frac{\pi \nu}{2}} H (-\tau)^{3/2} H_{i \nu}^{(2)}(-k \tau)\ , \\[4pt] 
\quad \bar\phi^*_k(\tau)&=\frac{\sqrt{\pi}}{2} e^{-\frac{\pi \nu}{2}} H (-\tau)^{3/2} H_{i \nu}^{(1)}(-k \tau) \ . 
\end{align}
\end{subequations}
Recall that $e^{\pi \nu}  H_{i \nu}^{(1)*}(-k \tau) =  H_{i \nu}^{(2)}(-k \tau)$, so that $\bar\phi_k(\tau)$ and $\bar\phi^*_k(\tau)$ are indeed related by complex conjugation, see \emph{e.g.}~\cite{Arkani-Hamed:2015bza} for a more detailed derivation and review. To simplify notation, we will always define $\bar \phi$ as the mode functions and drop the ``hat" on the field operators, $\hat \phi \to \phi$.

We re-rewrite $\bar \phi_{k}$ using the Mellin transform \cite{Watson:1944:TBF,Sleight:2019mgd,Premkumar:2021mlz}:
\begin{subequations}
\begin{align}
i \pi e^{-\frac{\pi \nu}{2}} H_{i \nu}^{(1)}(z) & =\int_{c-i \infty}^{c+i \infty} \frac{\mathrm{d} s}{2 \pi i} \Gamma\Big(s+\frac{i \nu}{2}\Big) \Gamma\Big(s-\frac{i \nu}{2}\Big)\Big(-\frac{i z}{2}\Big)^{-2 s} \ , \\[5pt]
-i \pi e^{\frac{\pi \nu}{2}} H_{i \nu}^{(2)}(z) & =\int_{c-i \infty}^{c+i \infty} \frac{\mathrm{d} s}{2 \pi i} \Gamma\Big(s+\frac{i \nu}{2}\Big) \Gamma\Big(s-\frac{i \nu}{2}\Big)\Big(\frac{i z}{2}\Big)^{-2 s} \ .
\end{align}
\end{subequations}
As we will now review, this formula compactly encodes the series representation of Hankel
function as a direct consequence of Cauchy's residue theorem.  The function $\Gamma(s\mp i\frac{\nu}{2})$ has poles at $s = \pm \frac{i\nu}{2}- n$, where $n=0,1,2,..$ is a non-negative integer; the residue of the $n^\text{th}$ pole is $(-1)^n/n!$.  We therefore have
\beq
\bar \phi_k(\tau) = i\frac{H(-\tau)^{3/2}}{2\sqrt{\pi}} \sum_{n=0}^\infty\frac{(-1)^n }{n!} \big( \Gamma(-i \nu -n)  \left(-\tfrac{i}{2} k \tau\right)^{i \nu +2n} +\Gamma(i \nu -n)  \left(-\tfrac{i}{2} k \tau\right)^{-i \nu +2n}\big) \ .
\label{eq:series}
\eeq
This is precisely a series expansion of the mode functions expanded around $k\tau = 0$. The leading order behavior in the limit $\tau \to 0$ is therefore just the first pole in the $\Gamma$ function
\beq
\bar \phi_k(\tau \to 0) \simeq i\frac{H(-\tau)^{3/2}}{2\sqrt{\pi}}\left( \Gamma(-i \nu) \left(-\tfrac{i}{2} k \tau\right)^{i \nu} + \Gamma(i \nu) \left(-\tfrac{i}{2} k \tau\right)^{-i \nu} \right) \ .
\label{eq:seriesLeading}
\eeq
At late times, the power spectrum of $\phi$ is therefore 
\begin{align}
\langle \phi(\k,\tau) \phi(\kp,\tau) \rangle &= \bar \phi^*_{k}(\tau \to 0)  \bar \phi_{k'}(\tau \to 0) (2\pi)^3 \delta(\k+\kp)  \notag\\[5pt]
&\simeq \frac{H^2 (-\tau)^3}{2\pi}\bigg( \Gamma(-i \nu)^2 \cosh(\tfrac{\pi \nu}{2} )\left(-\tfrac{k \tau}{2} \right)^{i 2 \nu} + \Gamma(i \nu)^2 \cosh(\tfrac{\pi \nu}{2} ) \left(-\tfrac{ k \tau}{2}\right)^{-i 2\nu}  \notag\\[3pt]
&\hspace{80pt}+\Gamma(-i \nu) \Gamma(i \nu) \cosh(\pi \nu) \bigg) (2\pi)^3 \delta(\k+\kp)\ .
\label{eq:twoPtSigmaLeading}
\end{align} 
Notice that this is the sum of power-laws in $k$, which suggests that each term arises from some kind of scaling operator. In particular, given a scale invariant theory, an operator in real space ${\cal O}(\x)$ with dimension $\dim$ will have a power spectrum in Fourier space of the form
\beq
\langle {\cal O}_\dim(\k\s) {\cal O}_\dim(\kp) \rangle \propto k^{-d + 2 \dim} (2\pi)^3 \delta(\k+\kp) \ ,
\eeq
where $d$ is the number of spacial dimensions (our interest here is the limit $d\to 3$). Therefore, the we can interpret the first two terms in \cref{eq:twoPtSigmaLeading} as resulting from a pair of operators ${\cal O}_\dim(\k\s)$ and ${\cal O}_{\bar\dim}(\k\s)$ with dimensions
\begin{subequations}
\begin{align}
\dim &= \frac{d}{2} + i \nu \ , \\[4pt]
\dimB & = \frac{d}{2} - i \nu = \dim^* = d -\dim\ .
\end{align}
\end{subequations}
The third term in \cref{eq:twoPtSigmaLeading} scales as $k^0$ and is the Fourier transform of a $\delta$-function in $\x$. We will refer to these terms as ``contact terms" although it is important that they are not localized in time.\footnote{This choice of terminology will be justified below in the context of SdSET, where time evolution is determined by dynamical RG flow.}

Our interpretation of the individual terms in the series expansion as arising from scaling ``operators" in not an accident: the isometeries of dS, $SO(4,1)$, act on the terms in this series at equal time as if they were part of a three-dimensional Euclidean CFT. It is for this reason that cosmological correlators in dS are subject to many powerful symmetry-based constraints. We will therefore use CFT terminology (dimensions, contact terms, etc) to describe these correlators because much of the CFT intuition will carry over.

For the calculations performed below, we will also need the commutator of $\phi$ at unequal times, $\tau \neq \tau'$. In the limit $k\tau, 
k'\tau' \to 0$, the commutator should scale as $k^0$, so that the commutator vanishes at spacelike separated points. Using the series expansion of $\bar\phi$ in \cref{eq:seriesLeading}, we find
\begin{align}
\langle [\phi(\k,\tau), \phi(\kp,\tau')] \rangle' &\simeq \frac{H^2(\tau \tau')^{3/2}}{2\pi}\Gamma(-i \nu) \Gamma(i \nu) \sinh(\pi \nu) \left( (-\tau)^{i\nu} (-\tau')^{-i\nu } -(-\tau')^{i\nu} (-\tau)^{-i\nu }\right) \notag\\[4pt]
&= \frac{H^2}{2\nu} \left( (-\tau)^{\dim} (-\tau')^{\bar \dim} -(-\tau')^{\dim} (-\tau)^{\bar \dim }\right) \ .
\label{eq:sigmaCom}
\end{align}
This result is consistent with conformal invariance. At separated points (in space), only operators of the same dimension have non-vanishing two point statistics, but this contact term is allowed because $\dim + \dimB = d$.

\subsection{Mass Mixing}

We can develop intuition that we can apply to loop corrections in an interacting theory below by studying the perturbative calculation of mixing between two massive fields.  Specifically, the model has one principal series field $\phi$ with dimensions $\dimPhi = \frac{3}{2} + i \nu_\phi$ and $\dimBPhi =3-\dimPhi$ that mixes with a second principal series field $\sigma$ with dimensions $\dimSig$ and $\dimBSig = 3- \dimSig$ via a mass mixing operator
\beq
{\cal H}_{\rm int} = \lambda H^2\phi \sigma \ .
\label{eq:HintMassMix}
\eeq
We can treat this mixing in the mass insertion approximation and calculate its impact on correlation functions of the operator $\phi$. The benefit of this example is that this mass mixing leads to a change in the mass of $\phi$, which changes the scaling dimension via $\nu_\phi = \sqrt{\frac{m^2}{H^2}-\frac{d^2}{4}}$.  We can then verify the result of the perturbative analysis by diagonalizing the mass matrix to calculate the dimensions exactly.

To see how this works, we will calculate the two point function of $\phi$ at ${\cal O}(\lambda^2)$ using in-in perturbation theory.  Following~\cite{Weinberg:2005vy}, this correlator can be written as
\beq
\langle {\cal Q}(t)\rangle=\left\langle\left[\bar{\text{T}} \exp \left(i \int_{-\infty}^t \d t\, H_{\rm int}(t) \right)\right] {\cal Q}^{\rm int}(t)\left[\text{T} \exp \left(-i \int_{-\infty}^t \d t\, H_{\rm int}(t)\right)\right]\right\rangle \ ,
\eeq
where ${\cal Q}(t)$ is some operator defined at a single time $t$ and ${\cal Q}^{\rm int}(t)$ is the same operator defined in terms of interaction picture fields. Expanding the time ordered expontentials allows one to write the in-in correlators in the commutator form
\begin{align}
\langle {\cal Q}(t) \rangle &=\sum_{N=0}^{\infty} i^N \int_{-\infty}^{t} \mathrm{d}t_N \int_{-\infty}^{t_N} \mathrm{~d} t_{N-1} \cdots \int_{-\infty}^{t_2} \mathrm{d}t_1 \notag\\[5pt]
& \hspace{35pt}
\times\left\langle\left[H_{\mathrm{int}}\left(t_1\right),\left[H_{\mathrm{int}}\left(t_2\right), \cdots\left[H_{\mathrm{int}}\left(t_N\right), \mathcal{Q}^{\mathrm{int}}(t)\right] \cdots\right]\right]\right\rangle \ .
\label{eq:commutator}
\end{align}
For the explicit calculations, we will introduce conformal time $\tau$, so that $\d t_i = a(\tau_i) \d\tau_i$, and then use the Mellin representation of the field operators.  Note that the commutator form of the in-in correlator is not compatible with the $i\epsilon$ prescription. In the Mellin-representation, all time integrals are defined by analytic continuation (like dimensional regularization) and therefore one can circumvent the need to impose the $i\epsilon$ prescription explicitly.

Calculating the correction to the power spectrum at ${\cal O}(\lambda^2)$ corresponds to the case $N=2$ and ${\cal Q} = \phi(\k,\tau) \phi(\kp,\tau)$. The leading correction in the superhorizon limit $-k\tau \ll 1$ takes the form (shown on the left in \cref{fig:dS_mix})
\begin{align}
&\langle\phi(\k,\tau)\phi(\kp,\tau)\rangle' =\label{eq:mixing_com} \\[3pt]
& -\lambda^2 H^4\int_{-\infty}^{\tau}\d\tau_2\s a(\tau_2)^4 \big[\phi(\k,\tau_2),\phi(\kp,\tau)\big]\int_{-\infty}^{\tau_2} \d\tau_1 a(\tau_1)^4 \langle\phi(\kp,\tau_1)\phi(\k,\tau)\rangle \big[\sigma(\k,\tau_1),\sigma (\kp,\tau_2)\big]\notag\\[3pt]
&-\lambda^2 H^4\int_{-\infty}^{\tau}\d\tau_2\s a(\tau_2)^4 \big[\phi(\k,\tau_2),\phi(\kp,\tau)\big]\int_{-\infty}^{\tau_2} \d \tau_1\s a(\tau_1)^4 \big[\phi(\kp,\tau_1),\phi(\k,\tau)\big]\langle\sigma(\k,\tau_1)\sigma(\kp,\tau_2)\rangle\ , \nonumber
\end{align}
where we have expanded the double commutator from \cref{eq:commutator}, and all terms inside the time integrals are evaluated at $\lambda=0$. 

This correction is just shift of the masses of $\sigma$ and $\phi$ after diagonalizing the mass matrix. As such, it will also shift the dimensions $\dimPhi  \to \dimPhi + \gamma_1$ and $\dimBPhi  \to \dimBPhi +\gamma_2$ where $\gamma_{1,2} = {\cal O}(\lambda^2)$. Expanding the free two-point function, we should then have 
\begin{align}
\langle \phi(\k,\tau)\phi(\kp,\tau)\rangle & \xrightarrow{-k \tau\rightarrow 0}\langle \phi(\k,\tau)\phi(\kp,\tau)\rangle_{\dimPhi}\left(1+ 2 \gamma_1 \log(-k \tau) +\ldots\right)\notag\\[4pt]
&\hspace{50pt}+\langle \phi(\k,\tau)\phi(\kp,\tau)\rangle_{\dimBPhi}(1+2\gamma_2\log(-k \tau) +\ldots) \ ,
\end{align}
where we defined
\begin{subequations}
\bea
\langle \phi(\k,\tau)\phi(\kp,\tau)\rangle'_{\dimPhi} &=&\frac{H^2 (-\tau)^3}{2\pi} \Gamma(-i \nu_\phi)^2 \cosh(\tfrac{\pi \nu_\phi}{2} )\left(-\tfrac{k \tau}{2} \right)^{i 2 \nu_\phi} \ ,  \\[6pt]
\langle \phi(\k,\tau)\phi(\kp,\tau)\rangle'_{\dimBPhi} &=&\frac{H^2 (-\tau)^3}{2\pi}\Gamma(i \nu_\phi)^2 \cosh(\tfrac{\pi \nu_\phi}{2} ) \left(-\tfrac{ k \tau}{2}\right)^{-i 2\nu_\phi} \ ,
\eea
\end{subequations}
the power law-correlators of the $\lambda = 0$ theory.
For the purpose of calculating $\gamma_{1,2}$ in perturbation theory, it is therefore sufficient to isolate the logarithmically enhanced terms. Since these are large when $\tau = 0$, we can determine these contributions from the region of integration where $-\tau_{1,2} k \ll 1$.

Using \cref{eq:sigmaCom} for both the $\phi$ and the $\sigma$ commutators in the superhorizon limit in the in-in formula, we have 
\begin{align}
  &\hspace{-15pt} -\lambda^2 H^4 \int_{-\infty}^{\tau}\d\tau_2\s a(\tau_2)^4 \big[\phi(\k,\tau_2),\phi(\kp,\tau)\big]\int_{-\infty}^{\tau_2} \d\tau_1\s a(\tau_1)^4 \langle\phi(\kp,\tau_1)\phi(\k,\tau)\rangle \big[\sigma(\k,\tau_1),\sigma (\kp,\tau_2)\big]\notag\\[3pt]
 =&-\lambda^2 \langle\phi(\kp,\tau)\phi(\k,\tau) \rangle_{\dimPhi} \frac{H^8}{2 \nu_\phi \nu_\sigma} \int_{-\infty}^{\tau}\d\tau_2\s a(\tau_2)^4 \left( (-\tau_2)^{\dimPhi} (-\tau)^{\dimBPhi} -(-\tau)^{\dimPhi} (-\tau_2)^{\dimBPhi }\right)\notag\\
 &\int_{-\infty}^{\tau_2} \d\tau_1\s a(\tau_1)^4  \left(\frac{\tau_1}{\tau}\right)^{\dimPhi} \left( (-\tau_1)^{\dimSig} (-\tau_2)^{\dimBSig} -(-\tau_2)^{\dimSig} (-\tau_1)^{\dimBSig }\right) - \bigg[ \dimPhi \leftrightarrow \dimBPhi \bigg] \notag\\[3pt]
 =&-\lambda^2\langle\phi(\kp,\tau)\phi(\k,\tau) \rangle_{\dimPhi} \frac{H^4}{2 \nu_\phi \nu_\sigma} \int_{-\infty}^{\tau}\d\tau_2 a(\tau_2)^4 \left( (-\tau_2)^{\dimPhi} (-\tau)^{\dimBPhi} -(-\tau)^{\dimPhi} (-\tau_2)^{\dimBPhi }\right)\notag\\
 &  \left(-\tau\right)^{-\dimPhi} (-\tau_2)^{\dimPhi+\dimSig + \dimBSig - 3} \left(\frac{1}{3- \dimPhi - \dimSig} - \frac{1}{3- \dimPhi - \dimBSig} \right) - \bigg[ \dimPhi \leftrightarrow \dimBPhi \bigg]\ ,
 \end{align}
where we defined the $k^{2\dimPhi -3}$ component of the power spectrum as $\langle\phi(\kp,\tau)\phi(\k,\tau) \rangle_{\dimPhi}$. The second term with $\dimBPhi$ in place of $\dimPhi$ appears with a minus sign to account for change of sign of the commutator after exchanging the two dimensions. In the last step we evaluated the $\tau$ integrals by analytic continuation from the region where they converge at $\tau_1 \to -\infty$. There is no particular subtlety required for $\tau_1$, since this choice also implements the $i\epsilon$ prescription which would have ensured convergence. The $\tau_2$ integral is more subtle because it contains a log-divergence. Since we are only interested in determining the coefficient of the log, the choice of regulator or scheme does not impact the result (to this order in perturbation theory). Therefore, we will simply analytically continue\footnote{As has been noted in previous work on SdSET, dimensional regularization is insufficient to regulate the log. Specifically, if we wrote $\dimPhi +\dimBPhi =d$ and analytically continued in the number of space dimensions consistently, one finds this integral diverges for all $d$. It is therefore essential that we are analytically continuing in $\dimPhi +\dimBPhi$ for fixed $d=3$.} in $\dimPhi +\dimBPhi $ and isolate the log in the limit $\dimPhi +\dimBPhi  \to d=3$. Evaluating the $\tau_2$ integral is this way, we find
 \begin{align}
 &\hspace{-20pt}-\langle\phi(\kp,\tau)\phi(\k,\tau) \rangle_{\dimPhi} \frac{\lambda^2}{2 \nu_\phi \nu_\sigma} (-\tau)^{\dimPhi+\dimBPhi - 3}\left(\frac{1}{3-2\dimPhi} - \frac{1}{3-\dimPhi-\dimBPhi} \right)  \notag \\[3pt]
 &\hspace{200pt} \times \left(\frac{1}{3- \dimPhi - \dimSig} - \frac{1}{3- \dimPhi - \dimBSig} \right) \notag \\[5pt]
 &\to -\langle\phi(\kp,\tau)\phi(\k,\tau) \rangle_{\dimPhi} \frac{\lambda^2}{2 \nu_\phi \nu_\sigma} \log(-k\tau) \left(\frac{1}{\dimBSig - \dimPhi} - \frac{1}{\dimSig - \dimPhi} \right) \ .
\end{align}
Additionally, we perform a similar calculation for the $k^{2\dimBPhi -3}$ component of the power spectrum  $\langle\phi(\kp,\tau)\phi(\k,\tau) \rangle_{\dimBPhi}$.  We then identify 
\begin{subequations}\label{eq:mix_dims}
\begin{align}
\gamma_1 &= -\frac{\lambda^2}{4 \nu_\phi \nu_\sigma}\left(\frac{1}{\dimBSig - \dimPhi} - \frac{1}{\dimSig - \dimPhi} \right) \ , \\[4pt]
\gamma_2 &=  \frac{\lambda^2}{4 \nu_\phi \nu_\sigma}\left(\frac{1}{\dimBSig - \dimBPhi} - \frac{1}{\dimSig - \dimBPhi} \right)\ . 
\end{align}
\end{subequations}
Using $\dimSig+ \dimBSig = 3$, $\dimPhi = 3/2 + i\nu_\phi$, and $\dimBPhi = 3/2-i\nu_\phi$, we see that $\gamma_1 = - \gamma_2$,
which is consistent with our expectation that the mass mixing operator leads to a shift in the mass of $\phi$.

Now we can compare this result with the direct diagonzalization of the mass matrix. We start again from the  the UV Lagrangian 
\beq
{\cal L} = -\frac{1}{2}\left(\partial_\mu \phi\s \partial^\mu\phi + m_\phi^2 \phi^2 \right)-\frac{1}{2}\left(\partial_\mu \sigma \partial^\mu\sigma + m_\sigma^2 \sigma^2 \right) - \lambda H^2 \phi \sigma  \ .
\eeq
Writing the mass term in matrix form, 
\beq
{\cal L} \supset - \frac{1}{2}\begin{pmatrix}
 \phi & \sigma 
\end{pmatrix}  \begin{pmatrix}
m_\phi^2 & \lambda H^2 \\
\lambda H^2 & m_\sigma^2
\end{pmatrix}  \begin{pmatrix}
 \phi \\ \sigma 
\end{pmatrix}\ ,
\eeq 
we can find the mass eigenvalues in terms of $\lambda$. Assuming $m^2_\sigma > m^2_\phi$ and $|m^2_\sigma - m_\phi^2| \gtrsim m_\phi^2$, the mass of $\phi$ is shift by
\beq
m_\phi^2 \to m_\phi^2 - \frac{\lambda^2 H^4}{|m_\sigma^2 -m_\phi^2|} + {\cal O}(\lambda^4)\ .
\eeq
Assuming $m_\phi^2 > 9 H^2/4$, we can expand the solution for $\dimPhi$ in $\lambda$ to find
\beq
\dimPhi + \gamma_1  = \frac{3}{2} + i \nu_\phi - \frac{i}{2 \nu_\phi}\frac{\lambda^2 H^2}{|m_\sigma^2 -m_\phi^2|}  \quad\to\quad \gamma_1 = - \frac{i}{2 \nu_\phi}\frac{\lambda^2 H^2}{|m_\sigma^2 -m_\phi^2|} \ .
\eeq
We can compare to our perturbative formula, \cref{eq:mix_dims}, using $\dimSig = \frac{3}{2} + i \nu_\sigma$, $\dimBSig = \frac{3}{2} - i \nu_\sigma$ and $\dimPhi=\frac{3}{2} + i \nu_\phi$ to find 
\beq
\gamma_1 = -\frac{\lambda^2}{4 \nu_\phi \nu_\sigma}\left(\frac{1}{-i\nu_\sigma  - i\nu_\phi} - \frac{1}{i\nu_\sigma - i\nu_\phi} \right) = - \frac{\lambda^2}{4 \nu_\phi \nu_\sigma} \frac{2 i \nu_\sigma}{\nu_\sigma^2 - \nu_\phi^2} = - i \frac{\lambda^2 H^2 }{4 \nu_\phi (m_\sigma^2-m_\phi^2)} \ .
\eeq
This confirms that the method of the perturbative calculation is correctly reproducing the expected result.

\subsection{One-loop Anomalous Dimensions From Mellin Space}
\label{sec:oneLoopFromMellin}
The one-loop correction to the power spectrum is equivalent to the mixing between a fundamental field $\phi$ with a composite operator $\sigma^2$. As a result, the calculation of the anomalous dimension of $\phi$ can be broken up into two components, the calculation of the Green's functions of $\sigma^2$ and the perturbative mixing between $\phi$ and $\sigma^2$. While this is mathematically identical to completing the one-loop calculation in a single step, we will present the results in this way to help isolate the important steps and to make closer contact with the SdSET results in the next section.

\subsubsection*{Composite Operators at One-loop}

In order to calculate the behavior of the composite operator $\sigma^2$, we need an approach to calculating loop integrals that respects the scale invariance that naively forbids the kinds of contact terms that are needed for mixing. It is for this reason that we will introduce the Mellin representation of the mode functions. The Mellin transform represents the full answer as an integral over power-laws. This is useful because the loop momentum intergal for individual power laws can be controlled with a scaleless regulator that preserves the symmetries of the theory in dS. The final (regulated) result then follows from performing the Mellin integrals (i.e.~the inverse Mellin transform).

The Mellin representation of the mode functions allows for a straightforward evaluation of the 
$[\sigma^n](\k,\tau)$ power spectrum, where $[\sigma^n](\k,\tau)$ is the composite operator associated with $\sigma^n(\x)$ in Fourier space. From Wick contractions, the two point function is given in terms of mode functions as
\begin{align}
\langle  \sigma^2(\k,\tau) \sigma^2(\kp,\tau') \rangle' = 2\int \frac{\d^3p}{(2 \pi)^3} \bar \sigma_{\p}(\tau) \bar \sigma_{\vec{k}-\p}(\tau)\bar \sigma^*_{-\p}(\tau') \bar \sigma^*_{-\vec{k} +\p}(\tau') \ .
\end{align}
Using our above expression, this is written in Mellin variables $s_1,..,s_4$ as
\begin{align}
\langle  \sigma^2(\k,\tau) \sigma^2(\kp,\tau') \rangle' &=  \frac{H^4(-\tau)^{3}(-\tau')^3}{8\pi^2 } \prod_{i=1}^4 \left(\int_{c-i \infty}^{c+i \infty} \frac{\mathrm{d} s_i}{2 \pi i}\Gamma\left(s_i+\frac{i \nu}{2}\right) \Gamma\left(s_i-\frac{i \nu}{2}\right)\right) \notag\\[3pt]
&\hspace{-12pt}\times \int \frac{\d^3 p}{(2 \pi)^3}(-\tfrac{i}{2}p \tau)^{-2 s_1} (-\tfrac{i}{2}|\k-\p| \tau)^{-2 s_2}(\tfrac{i}{2}p \tau')^{-2 s_3}(\tfrac{i}{2}|\p-\k| \tau')^{-2 s_4} \ .
\end{align}
As described above, we are interested in whether this integral generates a contact term in space, which would Fourier transform to a $k$-independent term, ${\cal O}(k^0)$. We can therefore expand the integrand in $p \gg k$, which amounts to evaluating the integrals with $k=0$:
\begin{align}
\langle  \sigma^2(\k,\tau) \sigma^2(\kp,\tau') \rangle' &\to  \frac{H^4(-\tau)^{3}(-\tau')^3}{8\pi^2 } \prod_{i=1}^4 \left(\int_{c-i \infty}^{c+i \infty} \frac{\mathrm{d} s_i}{2 \pi i}\Gamma\bigg(s_i+\frac{i \nu}{2}\bigg) \Gamma\left(s_i-\frac{i \nu}{2}\right)\right) \nonumber\\
&\hspace{12pt}\times \int \frac{\d^3 p}{(2 \pi)^3}(-\tfrac{i}{2} p \tau)^{-2 s_1-2 s_2} (\tfrac{i}{2}p \tau')^{-2 s_3-2 s_4}\ . \label{eq:mellin_loop}
\end{align}
We can evaluate the momentum integral as follows
\beq
\int \frac{\d^3 p}{(2 \pi)^3} p^{-2 S} = \frac{1}{2\pi^2}\lim_{p_0\to 0} \int^\infty_{p_0} \d p\, p^{2-2 S} =\frac{1}{2\pi^2} \lim_{p_0\to 0} \frac{p_0^{3-2S}}{3- 2 S} \simeq -\frac{i}{2 \pi}\delta\big(\tfrac{3}{2}- S\big)\ .
\eeq
The appearance of the $\delta$-function in the final step is conventional in both AdS~\cite{Fitzpatrick:2011hu} and dS~\cite{Sleight:2019mgd,Premkumar:2021mlz}. However, it is ambiguous given that we will be integrating over $s_i$ contours in the complex plane. We should understand the $\delta$-function to mean that we will evaluate one of the $s_i$ integrals by closing the contour around the simple pole at $2S=3$ using the residue formula. The residue of this pole is independent of the limit $p_0 \to 0$. In principle, we should worry how we close the contour around this pole (particularly the contour at infinity). However, like all scaleless regulators, we define the limit $p_0 \to 0$ to vanish for all power law divergences by analytic continuation. As we take this limit first, we are only left with the contribution from this pole, which effectively acts like a $\delta$ function on the contour of integration.

With this interpretation of the $\delta$-function, the correlator after performing the loop integral becomes
\begin{align}\label{eq:delta_int}
\langle  \sigma^2(\k,\tau) \sigma^2(\kp,\tau') \rangle' \to&  -i \frac{H^4(-\tau)^{3}(-\tau')^3}{16\pi^3 } \prod_{i=1}^4 \left(\int_{c-i \infty}^{c+i \infty} \frac{\mathrm{d} s_i}{2 \pi i}\Gamma\bigg(s_i+\frac{i \nu}{2}\bigg) \Gamma\bigg(s_i-\frac{i \nu}{2}\right)\bigg) \nonumber\\[4pt]
&\hspace{12pt}\times\delta\left(\tfrac{3}{2}- s_1-s_2-s_3-s_4\right)(-\tfrac{i}{2}  \tau)^{-2 s_1-2 s_2} (\tfrac{i}{2} \tau')^{-2 s_3-2 s_4} \ .
\end{align}
We can now evaluate one of the Mellin integrals using the $\delta$-function first, and then evaluate the remaining three Mellin integrals using the residue theorem. Evaluating the integral over $s_4$ using the delta function yields:
\begin{align}
\langle  \sigma^2(\k,\tau) \sigma^2(\kp,\tau') \rangle' \to&  - \frac{H^4(-\tau)^{3}(-\tau')^3}{32\pi^4 } \prod_{i=1}^3 \left(\int_{c-i \infty}^{c+i \infty} \frac{\mathrm{d} s_i}{2 \pi i}\,\Gamma\bigg(s_i+\frac{i \nu}{2}\bigg) \Gamma\bigg(s_i-\frac{i \nu}{2}\right)\bigg) \nonumber\\[4pt]
&\hspace{12pt}\times\Gamma\bigg(s_4+\frac{i\nu}{2}\bigg)\Gamma\bigg(s_4-\frac{i\nu}{2}\bigg)(-\tfrac{i}{2}  \tau)^{-2 s_1-2 s_2} (\tfrac{i}{2} \tau')^{2s_1+2s_2-3} \ ,
\end{align}
where
\beq
s_4 = \frac{3}{2}  -s_1-s_2-s_3\ .
\eeq
We evaluate the remaining integrals over $s_1,s_2,s_3$ using the residue theorem by closing all the contours in the lower-half plane. Note that the exponents of $\tau$ and $\tau'$ depend on $s_1$ and $s_2$; the contour integrals will only converge in the limits $\text{Re}(s_1),\text{Re}(s_2)\to -\infty$ if $(-\tau) > (-\tau')$. 

Following the same procedure as we did for the series expansion of $\bar \phi$ given in \cref{eq:series}, we now evaluate the integrals $s_1,s_2,s_3$ over the poles of the $\Gamma$ functions, which yields a sum over integers $n_1,n_2,n_3$:
\begin{align}\label{eq:triple_sum}
\langle  \sigma^2(\k,\tau) \sigma^2(\kp,\tau') \rangle' \to&  - \frac{H^4}{32\pi^4 } \sum_{\pm_1,\pm_2,\pm_3 } \sum_{n_1,n_2,n_3=0}^\infty  \Gamma\left(s_4+\frac{i \nu}{2}\right) \Gamma\left(s_4-\frac{i \nu}{2}\right) \notag\\[6pt]
&\hspace{12pt}\times\frac{(-1)^{n_1+n_2+n_3} }{n_1!n_2! n_3!}\Gamma(\mp_1i \nu -n_1) \Gamma(\mp_2i \nu -n_2)\Gamma(\mp_3i \nu -n_3) \notag \\[6pt]
&\hspace{12pt}\times\left(-\tfrac{i}{2} \tau\right)^{3+(\pm_1 1\pm_2 1)i \nu +2(n_1+n_2)}\left(\tfrac{i}{2} \tau'\right)^{-(\pm_1 1\pm_2 1)i \nu -2(n_1+n_2)} \ ,
\end{align}
where
\beq\label{eq:s4_vals}
s_4 = \frac{3}{2} + i (\pm_1 1\pm_2 1\pm_31) \frac{\nu}{2} +n_1+n_2+n_3 \ .
\eeq
Note that every term in the sum has the form 
\beq
\langle  \sigma^2(\k,\tau) \sigma^2(\kp,\tau') \rangle' \propto (-\tau)^{\Delta} (-\tau')^{3-\Delta} \ ,
\eeq
where 
\beq
\Delta = 3+(\pm_1 1\pm_2 1)i \nu +2(n_1+n_2)  = \Delta_{n_1} + \Delta_{n_2}
\eeq
is the scaling dimension we get from adding two terms in the series expansion of $\sigma$. This is the expected set of scaling dimensions for the composite operators composed of $\sigma^2$ and derivatives in a free-theory.

So far, the result is not written as a sum of dimensions, as there are many choices of $s_{1,2,3}$ that given the same $\Delta$. As a concrete example, we can focus on the terms with $n_1=n_2=0$, $\pm_1= +$, and $\pm_2 = -$. This fixes the powers in $\tau$ and $\tau'$ to $\Delta =3$ and $\bar \Delta =3-\Delta =0$ respectively. Even with these powers fixed, we must still perform the sum over $n_3$: 
\begin{align}
\langle  \sigma^2(\k,\tau) \sigma^2(\kp,\tau') \rangle' \supset&  - \bigg(\frac{i}{2}\bigg)^3 \frac{H^4(-\tau)^{3}}{64\pi^4 } \sum_{\pm_3 } \sum_{n_3=0}^\infty \frac{(-1)^{n_3} }{n_3!} \Gamma(i \nu) \Gamma(-i \nu )\Gamma(\pm_3i \nu -n_3) \notag \\[4pt]
&\hspace{12pt}\times\Gamma\left(\frac{3}{2}+n_3+(\pm_3+1)\frac{i \nu}{2}\right) \Gamma\left(\frac{3}{2} +n_3+(\pm_3-1)\frac{i \nu}{2}\right) \ .
\end{align}
We see that reducing our answer to a sum over composite operators requires some further simplifications.

Fortunately, the sum over all the residues of $s_{1,2,3}$ can be reduced to a single sum. Assuming $(-\tau) < (-\tau')$, we can rewrite \cref{eq:triple_sum} as a single sum
\begin{align}\label{eq:single_sum}
\langle  \sigma^2(\k,\tau) \sigma^2(\kp,\tau') \rangle' &\to \sum_n a_n(\nu) (-\tau)^{2 i\nu+2n+3}(-\tau')^{-2i\nu -2n} +(\nu \rightarrow -\nu)\nonumber\\[4pt] 
&\hspace{16pt}+\sum_n c_n(\nu) (-\tau)^{2n+3}(-\tau')^{ -2n} \ ,
\end{align}
where the series coefficients $a_n(\nu)$ and $c_n(\nu)$ are
\begin{subequations}
\begin{align}
a_n(\nu)&=-i\frac{H^4\left(1+\tanh(2\pi \nu)\right)}{16\pi^2 }\frac{\Gamma(\frac{1}{2}+n+i\nu)\Gamma(\frac{3}{2}+n+i\nu)\Gamma(-1-n-i\nu)\Gamma(-n-i\nu)}{ n!\Gamma(-\frac{1}{2} - n) \Gamma(1 + n +2i \nu) \Gamma(-\frac{1}{2} - n - 2 i \nu)}\ , \\
c_n(\nu)&= i\frac{H^4\pi \text{csch}(2\pi \nu)^2}{n!(n+1)!4} \frac{\Gamma(\frac{1}{2}+n)\Gamma(\frac{3}{2}+n)}{\Gamma(1+n+i\nu)\Gamma(1+n-i\nu)\Gamma(-\frac{1}{2}-n-i\nu)\Gamma(-\frac{1}{2}-n+i\nu)} \ .
\end{align}%
\label{eq:coeffsSinglesum}%
\end{subequations}%
The details of how to turn the triple sum in \cref{eq:triple_sum} into a single sum can be found in \cref{app:sum}. It is useful to define $\Delta_{j=0,\pm,n}= 2i j \nu + 2n+3$, $\bar \Delta_{j=0,\pm,n} = 3 -\Delta_{j=0,\pm,n}$, and $C(\Delta_{j=\pm,n})= a_n(\pm\nu)$ and $C_{0,n} = c_n(\nu)$ so that   
\beq\label{eq:comp_2pt_jn}
\langle  \sigma^2(\k,\tau) \sigma^2(\kp,\tau') \rangle' \to  \sum_{j,n} C(\Delta_{j,n})(-\tau)^{\Delta_{j,n}}(-\tau')^{\bar \Delta_{j,n}} \ ,
\eeq
where to simplify the notation, we have defined $\sum_{j,n} \equiv \sum_{j=0,\pm} \sum_{n=0}^\infty$.

We see that $\tau$ is associated with $\Delta_{j,n}$, while $\tau'$ is associated with $\bar \Delta_{j,n}$. This is a consequence of the fact that when performing the triple sum, we had to assume $(-\tau) <(-\tau')$ in order for the sum to converge. This is important, as our in-in calculations are not time-ordered. For example, we cannot determine the commutator simply by exchanging $\tau \leftrightarrow \tau'$. 

We therefore must perform another calculation to determine $\langle  \sigma^2(\k,\tau) \sigma^2(\kp,\tau') \rangle'$ when $(-\tau) > (-\tau')$, in that we have to evaluate \cref{eq:delta_int} in a slightly different order. In this case, we evaluate the integrals over $s_2,s_3,s_4$ using the residue theorem and then evaluate the integral over $s_1$ using the delta function, to get
\begin{align}
\langle  \sigma^2(\k,\tau) \sigma^2(\kp,\tau') \rangle' \to&  - \frac{H^4(-\tau)^{3}(-\tau')^3}{32\pi^4 } \sum_{\pm_2,\pm_3,\pm_4 } \sum_{n_2,n_3,n_4=0}^\infty \Gamma\left(s_1+\frac{i \nu}{2}\right) \Gamma\left(s_1-\frac{i \nu}{2}\right) \notag\\[6pt]
&\hspace{12pt}\times\frac{(-1)^{n_2+n_3+n_4} }{n_2!n_3! n_4!} \Gamma(\mp_2i \nu -n_2) \Gamma(\mp_3i \nu -n_3)\Gamma(\mp_4i \nu -n_4) \notag \\[6pt]
&\hspace{12pt}\times\left(-\tfrac{i}{2} \tau\right)^{-(\pm_3 1\pm_4 1)i \nu -2(n_3+n_4)-3}\left(\tfrac{i}{2} \tau'\right)^{(\pm_3 1\pm_4 1)i \nu +2(n_3+n_4)} \ ,
\end{align}
where
\beq
s_1 = \frac{3}{2} + i (\pm_2 1\pm_3 1\pm_4 1) \frac{\nu}{2} +n_2+n_3+n_4 \ .
\eeq
This is related to the above result by replacing $n_3\to n_1, n_4\to n_2, n_2\to n_3, n_1\to n_4$, so we have
\begin{align}
\langle  \sigma^2(\k,\tau) \sigma^2(\kp,\tau') \rangle' &\to    \sum_n -e^{-4\pi \nu}a_n(\nu) (-\tau)^{-2 i\nu-2n}(-\tau')^{2i\nu +2n+3} -(\nu \rightarrow -\nu)\nonumber\\[4pt] 
&\hspace{16pt}+\sum_n -c_n(\nu) (-\tau)^{-2n}(-\tau')^{2n+3}\ .
\end{align}
Notice that the exponent of $\tau'$ is now the one with the largest real part instead of $\tau$. Using the fact that $a_n(-\nu)= -a_n(\nu)^* e^{-4\pi \nu} $ and $-c_n(\nu)= c_n(\nu)^*$, we can rewrite the above as
\begin{align}
\langle  \sigma^2(\k,\tau) \sigma^2(\kp,\tau') \rangle' &\to    \sum_n (a_n(-\nu))^* (-\tau)^{-2 i\nu-2n}(-\tau')^{2i\nu +2n+3} +(\nu \rightarrow -\nu)\nonumber\\[4pt]
&\hspace{16pt}+\sum_n c_n(\nu)^* (-\tau)^{-2n}(-\tau')^{2n+3} \ .
\end{align}
We can write this more compactly as  
\begin{align}
\langle  \sigma^2(\k,\tau) \sigma^2(\kp,\tau') \rangle' &\to  \theta(\tau -\tau') \sum_{j,n} C(\Delta_{j,n})(-\tau)^{\Delta_{j,n}}(-\tau')^{\bar \Delta_{j,n}} \notag\\[4pt]
&\hspace{16pt}+\theta(\tau'-\tau) \sum_{j,n}  C(\Delta_{j,n})^*(-\tau)^{\bar\Delta_{-j,n}}(-\tau')^{\Delta_{-j,n}} \ ,
\end{align}
such that the commutator takes the form
\begin{align}\label{eq:sig2_com}
\langle  [\sigma^2(\k,\tau) &,\sigma^2(\kp,\tau') ]\rangle' \to  \\[2pt]
&  \theta(\tau -\tau')\sum_{j,n} C(\Delta_{j,n}) (-\tau)^{\Delta_{j,n}}(-\tau')^{\bar \Delta_{j,n}}-C(\Delta_{j,n})^* (-\tau)^{\Delta_{-j,n}}(-\tau')^{\bar \Delta_{-j,n}}\notag\\[4pt]
+\,&\theta(\tau'-\tau) \sum_{j,n}C(\Delta_{j,n})^*(-\tau)^{\bar\Delta_{-j,n}}(-\tau')^{\Delta_{-j,n}}-
C(\Delta_{j,n})(-\tau)^{\bar\Delta_{j,n}}(-\tau')^{\Delta_{j,n}} \ . \notag
\end{align}
These results will be used to compute the 1-loop correction to the anomalous dimensions of a principal series field, which we turn to next.

\subsubsection*{Anomalous Dimensions}
\begin{figure}[t!]
    \centering
    \includegraphics[width=\textwidth]{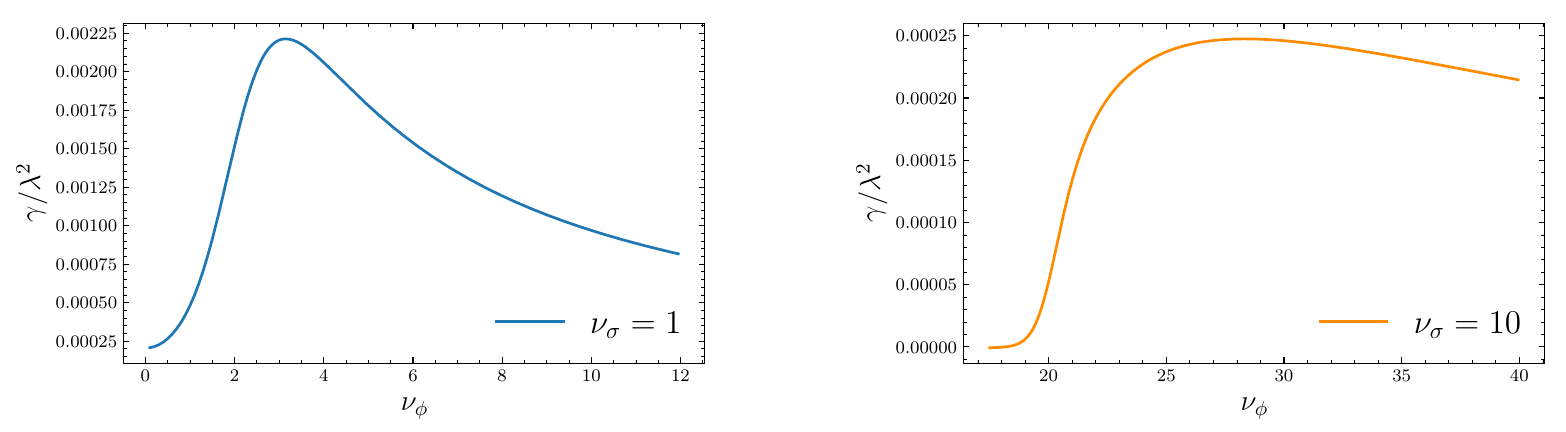}
    \caption{The real part of the anomalous dimension of $\phi$ as a function of $\nu_\phi$, holding fixed $\nu_\sigma=1$ (left) and $\nu_\sigma=10$ (right). Both panels are computed numerically from the sum in \cref{eq:anom_final} up to $n=4000$. We see the tendency for a larger anomalous dimension when $\nu_\phi \simeq 2 \nu_\sigma$.}
    \label{fig:fix_loop}
\end{figure}

\begin{figure}[h!]
    \centering
    \vspace{25pt}
    \includegraphics[width=0.55\textwidth]{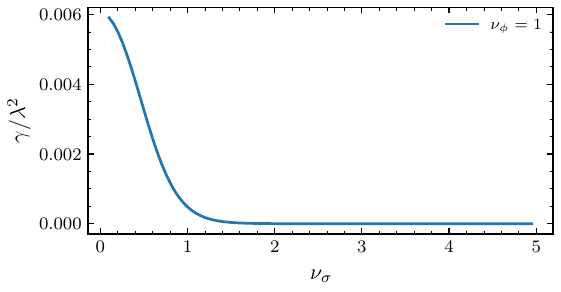}
    \caption{The real part of the anomalous dimension of $\phi$ as a function of $\nu_\sigma$, holding fixed $\nu_\phi=1$. The result is calculated numerically from \cref{eq:anom_final} up to $n=2000$. The dominant feature when varying the mass of the field in the loop is to suppress the amplitude (at fixed $\nu_\sigma$).}
    \label{fig:fix_ex}
\end{figure}

We now show how to compute the anomalous dimension of a principal series field $\phi$ due to a loop of another principal series field $\sigma$.
The interaction between these two fields is given by
\beq
{\cal H}_{\rm int} = \lambda H \phi \sigma^2 \ .
\label{eq:Hintcubic}
\eeq
The one-loop diagram (shown on the right in \cref{fig:dS_mix}) that corrects the $\phi$ power spectrum can be expressed following the same logic that led to \cref{eq:mixing_com}:
\begin{align}
&\hspace{0pt}\langle\phi(\k,\tau)\phi(\kp,\tau)\rangle' = \nonumber\\[4pt]
& -\lambda^2 H^2\int_{-\infty}^{\tau}\d\tau_2 a(\tau_2)^4 \big[\phi(\k,\tau_2),\phi(\kp,\tau)\big]\int_{-\infty}^{\tau_2} \d\tau_1 a(\tau_1)^4 \langle\phi(\kp,\tau_1)\phi(\k,\tau)\rangle \langle\big[\sigma^2(\k,\tau_1),\sigma^2 (\kp,\tau_2)\big]\rangle\notag\\[4pt]
&-\lambda^2 H^2\int_{-\infty}^{\tau}\d\tau_2 a(\tau_2)^4 \big[\phi(\k,\tau_2),\phi(\kp,\tau)\big]\int_{-\infty}^{\tau_2} \d \tau_1 a(\tau_1)^4 \big[\phi(\kp,\tau_1),\phi(\k,\tau)\big]\langle\sigma^2(\k,\tau_1)\sigma^2(\kp,\tau_2)\rangle\ , \label{eq:s2_com}
\end{align}
Using \cref{eq:sigmaCom} for the $\phi$ commutator and \cref{eq:sig2_com} for the $\sigma^2$ commutator, we have 
\begin{align}
 &\hspace{-10pt} -\lambda^2 H^2 \int_{-\infty}^{\tau}\d\tau_2 a(\tau_2)^4 \big[\phi(\k,\tau_2),\phi(\kp,\tau)\big]\int_{-\infty}^{\tau_2} \d\tau_1 a(\tau_1)^4 \langle\phi(\kp,\tau_1)\phi(\k,\tau)\rangle \langle\big[\sigma^2(\k,\tau_1),\sigma^2 (\kp,\tau_2)\big]\rangle\notag\\[4pt]
 &\to -\lambda^2 \langle\phi(\kp,\tau)\phi(\k,\tau) \rangle_{\Delta_{\phi}}\frac{H^4}{2\nu_{\phi}}\int_{-\infty}^{\tau}\d \tau_2 a(\tau_2)^4  \left( (-\tau_2)^{\Delta_\phi} (-\tau)^{\bar{\Delta}_{\phi}} -(-\tau)^{\Delta_\phi} (-\tau_2)^{\bar{\Delta}_{\phi} }\right)\notag\\[4pt]
 &\hspace{10pt}\times \sum_{j,n}\int \d\tau_1 a(\tau_1)^4 \left(\frac{\tau_1}{\tau}\right)^{\Delta_{\phi}}
\left( C(\Delta_{j,n})^*(-\tau_1)^{\bar\Delta_{-j,n}}(-\tau_2)^{\Delta_{-j,n}}-
C(\Delta_{j,n})(-\tau_1)^{\bar\Delta_{j,n}}(-\tau_2)^{\Delta_{j,n}}\right)\notag\\[4pt]
&\to - \lambda^2 \langle\phi(\kp,\tau)\phi(\k,\tau) \rangle_{\Delta_{\phi}}\frac{\log(- k\tau)}{2 H^4 \nu_{\phi}}
\left(\sum_{j,n}\frac{C(\Delta_{j,n})^*}{3-\Delta_\phi-\bar{\Delta}_{-j,n}}-\frac{C(\Delta_{j,n})}{3-\Delta_\phi-\bar{\Delta}_{j,n}}\right) \ . \label{eq:anom_step1}
\end{align}
The anomalous dimensions are therefore 
\begin{subequations}\label{eq:anom_final}
\begin{align}
\gamma_{\Delta_\phi}&=-\lambda^2\frac{1}{4H^4 \nu_\phi}\left(\sum_{j,n}\frac{C(\Delta_{j,n})^*}{3-\Delta_\phi-\bar{\Delta}_{-j,n}}-\frac{C(\Delta_{j,n})}{3-\Delta_\phi-\bar{\Delta}_{j,n}}\right)\ ,\\[4pt]
\gamma_{\bar{\Delta}_\phi}&=\lambda^2\frac{1}{4H^4\nu_\phi}\left(\sum_{j,n}\frac{C(\Delta_{j,n})^*}{3-\bar{\Delta}_\phi-\bar{\Delta}_{-j,n}}-\frac{C(\Delta_{j,n})}{3-\bar{\Delta}_\phi-\bar{\Delta}_{j,n}}\right)\ .
\end{align}
\end{subequations}
The sums can be calculated numerically and the results are shown in Figures~\ref{fig:fix_loop} and~\ref{fig:fix_ex}.


If the external field $\phi$ is a principal series field, these expressions imply $\gamma_{\Delta_\phi}^*=\gamma_{\bar{\Delta}_\phi}$, which is consistent with the reality condition of $\phi$. On the other hand, if the external field $\phi$ is a complementary series field, then both $\gamma_{\Delta_\phi}$ and $\gamma_{\bar{\Delta}_\phi}$ are purely real and $\gamma_{\Delta_\phi} \neq -\gamma_{\bar{\Delta}_\phi}$, which implies that the effect of the loop is not a simple shift in the mass of $\phi$.

The numerically computed dimensions illustrate two additional features that are visible in these expressions. First, varying $\nu_\phi$ at fixed $\nu_\phi$ leaves the coefficients $C(\Delta_{j,n})$ fixed. As a result the $\nu_\phi$ dependence is driven by the pole inside the sum and the overall $\nu_\phi^{-1}$. The result is enhanced when $\nu_\phi \simeq 2 \nu_\sigma$ as seen in \cref{fig:fix_loop}. In contrast, increasing $\nu_\sigma$ decreases $C(\Delta_{j,n})$ and thus leads to a large suppression of $\gamma$, as seen in \cref{fig:fix_ex}.

Now consider the regime where the mass of the external field is much larger then the mass of the fields in the loop, \emph{i.e.}, $\nu_{\phi} \gg \nu$, then to leading order
\begin{subequations}
\begin{align}
\gamma_{\Delta_\phi} &\simeq -\lambda^2\frac{1}{4H^4 \nu_\phi}\left(\sum_{j,n}\frac{3/2+2n+i\mu}{(3/2+2n)^2+\mu^2}\bigg( C(\Delta_{j,n})^*-C(\Delta_{j,n})\bigg)\right) \ ,\\[4pt]
\gamma_{\bar{\Delta}_\phi} &\simeq \lambda^2\frac{1}{4H^4 \nu_\phi}\left(\sum_{j,n}\frac{3/2+2n-i\mu}{(3/2+2n)^2+\mu^2}\bigg( C(\Delta_{j,n})^*-C(\Delta_{j,n})\bigg)\right) \ .
\end{align}
\end{subequations}
In this limit, it is straightforward to check that $\gamma_{\dimPhi} =\gamma_{\dimBPhi} > 0$  is guaranteed term by term if ${\rm Im} C(\Delta_{j,n}) < 0$.

Overall, we have reproduced the expected results for the the anomalous dimensions of principal series fields by direct calculation at one-loop. The final results are qualitatively similar to the perturbative mixing of two fields, but is extended to the sum over dimensions that define the composite operator $\sigma^2$. This will be useful in providing physical intuition for how anomalous dimensions arise more generally.

\subsection{Relation to the K\"all\'en-Lehmann Representation}
The above anomalous dimensions have been derived previously using the K\"all\'en-Lehmann representation~\cite{Bros:2010rku,Marolf:2010zp} (see also~\cite{Loparco:2023rug,DiPietro:2023inn,Loparco:2023akg,Sleight:2021plv}). We would like to establish that the Mellin calculation produces identical results. This is important because the Mellin approach will naturally generalize to inflationary calculations, while the identification of a complete basis of states (required for the K\"all\'en-Lehmann approach) beyond the limit of a fixed dS background is more challenging.

The K\"all\'en-Lehmann representation of correlators in de Sitter is defined by inserting a complete set of states:
\beq
1=|\Omega\rangle\langle\Omega|+\sum_{\ell} \int \frac{\d \Delta}{2 \pi i} \frac{1}{N(\Delta, \ell)} \int \frac{\d^d k}{(2 \pi)^d}| \Delta, k\rangle_{\mu_1 \ldots \mu_{\ell}}{ }^{\mu_1 \ldots \mu_{\ell}}\langle\Delta, k|+\ldots \ ,
\eeq
where $N(\Delta,\ell)$ is a normalization factor and $\Delta$, $\k$, and $\mu_{1,..\ell}$ label the states in terms of eigenvalues under dilations, translations, and rotations respectively. For principal series operators, the integral over $\Delta$ is evaluated along the line $\Delta= 3/2+i \nu'$ for $\nu' \in (-\infty,\infty)$. On general grounds, one can write 
\beq\label{eq:KL}
\langle\Omega|\O(\tau, \x) \O(\tau^{\prime}, \vec{x}^{\hspace{1pt}\prime})| \Omega\rangle=\int_{\mathbb{D}} \frac{\d \nu'}{2 \pi} \rho_\phi\left(\frac{d}{2}+i \nu'\right) G(\xi ; \nu') \ ,
\eeq
where $\rho_\O(\Delta)$ is the spectral density, $\xi$ is the geodesic distance between the operators, and $G(\xi ; \nu')$ is the Green's function that satisfies
\beq
\nabla^2 G(\xi,\nu') + \Delta(3-\Delta)H^2 G(\xi,\nu') =\delta^4(\x,\tau) \ .
\eeq
Along the contour of integration, this is just the Green's functions of a free principal series field. It will be important that this definition can be analytically continued to $\Delta$ off the principal series line. 

As above, we will Fourier transform the comoving distance $\vec{x} \to \vec{k}$. The Green's function then takes the form
\beq
G(\k,\kp,\tau,\tau';\nu') =\frac{H^2 \pi}{4} (-\tau)^{\frac{3}{2}} (-\tau')^{\frac{3}{2}} H_{i\nu'}^{(1)}(-k \tau)H_{i\nu'}^{(2)}(-k \tau') (2\pi)^3 \delta(\k+\kp) \ .
\eeq
 At late times $\tau, \tau^\prime \rightarrow 0$, we can apply \cref{eq:series} to see that the Green's function includes a contact term of the form
\beq\label{eq:contact}
G(\k,\tau,\tau^\prime;\nu') \xrightarrow{\text{contact}}\frac{H^2}{4 \nu' \sinh(\pi \nu')}\left[e^{-\pi \nu'}(-\tau)^{d-\Delta}(-\tau^{\prime})^{\Delta}+e^{\pi \nu'}(-\tau)^{\Delta}(-\tau^{\prime})^{d-\Delta}\right] \ .
\eeq 
We should note that this is not the causal Green's function (commutator) and therefore there are non-contact contributions that do not vanish but we can anticipate will not be required for matching our calculation in the previous subsection.

For a principal series scalar $\sigma$ with dimension $\Delta_\sigma$, the density of states associated with the operator $\O =\sigma^2$ is given by~\cite{Hogervorst:2021uvp} 
\begin{align}
\hspace{-12pt}\rho_{\sigma^2}(\Delta) & =\frac{H^2\nu' \sinh (\pi \nu')}{2^3 \pi^{\frac{d}{2}+2} \Gamma\left(\frac{d}{2}\right)} \frac{\Gamma^2\left(\frac{\Delta}{2}\right)\, \Gamma^2\left(\frac{d-\Delta}{2}\right)}{\Gamma(\Delta) \Gamma(d-\Delta)}\Gamma\left(\frac{2 \Delta_\sigma+\Delta-d}{2}\right) \Gamma\left(\frac{2 \Delta_\sigma-\Delta}{2}\right) \notag \\[4pt]
&\hspace{12pt} \times  \Gamma\left(\frac{d-2 \Delta_\sigma+\Delta}{2}\right) \Gamma\left(\frac{2 d-2 \Delta_\sigma-\Delta}{2}\right) \ .
\end{align}
An important feature of this formula is that the poles of $\rho_{\sigma^2}(\Delta)$ in the complex plane correspond to the dimensions of the individual scaling operators associated with $\sigma^2$.
The spectral density $\rho_{\sigma^2}(\Delta)$ has three classes of single poles in the right half plane: 
\beq
\Delta=2 \Delta_\sigma+2n, \qquad \Delta=2\left(d-\Delta_\sigma\right)+2n, \qquad \text { and } \quad \Delta=d+2n, \quad n \in \mathbb{N} \ .
\eeq
One interpretation of the Green's function $G(\k,\kp,\tau,\tau';\nu')$ in \cref{eq:KL} is that it represents the propagator of a auxiliary field $\sigma_\nu$ with dimension $\Delta = \frac{3}{2} + i\nu'$.
We can therefore find the contact term contributions to $\langle \sigma^2 \sigma^2\rangle$ by replacing propagator of the auxiliary field inside the spectral integral with \cref{eq:contact}.
 \begin{align}
 \langle \sigma^2(\k,\tau)\sigma^2(\kp,\tau^\prime)\rangle &\rightarrow \int_{\frac{3}{2}-i\infty}^{\frac{3}{2}+i\infty} \frac{\d\Delta}{2\pi i} \frac{H^2 \rho_{\sigma^2(\Delta)}}{4 \nu' \sinh(\pi \nu')}\notag\\[4pt]
 &\hspace{60pt}\times\left[e^{-\pi \nu'}(-\tau)^{d-\Delta}(-\tau^{\prime})^{\Delta}+e^{\pi \nu'}(-\tau)^{\Delta}(-\tau^{\prime})^{d-\Delta})\right] \ .
 \end{align}
Since the two terms inside the square bracket are related by the so-called ``shadow transformation'' $\nu'\to -\nu'$, both terms give the same contribution to the integral, so we only need to evaluate one of them. 

One way to evaluate this integral is by closing the integration contour in the right-half plane, and then apply the residue theorem. Notice that the exponents of $\tau$ and $\tau'$ depend on $\Delta$ in the form $(\frac{\tau}{\tau'})^\Delta$ or  $(\frac{\tau'}{\tau})^\Delta$; therefore regardless of the time ordering, we cannot keep both terms as one of them will always diverge when $\text{Re}(\Delta)\to +\infty$. For the case $\tau> \tau'$, we need to drop the first term. Similarly for the case $\tau< \tau'$, dropping the second term yields
 \begin{align}
\hspace{-10pt} \langle \sigma^2(\k,\tau)\sigma^2(\kp,\tau')\rangle \rightarrow -\theta\left(\tau-\tau'\right)
 &\sum_{\Delta_i} {\rm Res}_{\Delta=\Delta_i} \rho_{\sigma^2}(\Delta)   \frac{H^2 e^{\pi \nu'}}{2\nu' \sinh(\pi \nu')}(-\tau)^{\Delta}(-\tau')^{d-\Delta}\notag\\
-\theta(\tau'-\tau) &\sum_{\Delta_i} {\rm Res}_{\Delta=\Delta_i} \rho_{\sigma^2}(\Delta)   \frac{H^2 e^{-\pi \nu'}}{2\nu' \sinh(\pi \nu')}(-\tau)^{d-\Delta}(-\tau')^{\Delta}\ .
\end{align}
The contact term becomes again a series sum
\begin{align}
&\langle  \sigma^2(\k,\tau) \sigma^2(\kp,\tau') \rangle' \xrightarrow{\text{contact}} \notag\\[4pt]
&\hspace{5pt} \theta\left(\tau-\tau'\right)\Big(\sum_n\left( \tilde{a}_n(\nu) (-\tau)^{2 i\nu+2n+3}(-\tau')^{-2i\nu -2n} +(\nu\rightarrow -\nu)\right)
+ \tilde {c}_n(\nu) (-\tau)^{2n+3}(-\tau')^{-2n}\Big) \notag\\[4pt]
&\hspace{12pt}+\theta(\tau'-\tau)\Big(\sum_n \left(\tilde{a}_n(-\nu)^*(-\tau')^{2 i\nu+2n+3}(-\tau)^{-2i\nu -2n} +(\nu\rightarrow -\nu)\right) \notag\\[3pt]
&\hspace{104pt}+\tilde {c}_n(\nu)^* (-\tau')^{2n+3}(-\tau)^{-2n}\Big) \ ,
\end{align}
where $\tilde{a}_n(\nu)$ and $\tilde{c}_n(\nu)$ are given by 
\begin{subequations}
\begin{align}
\tilde{a}_n(\nu)&= \frac{i(-1)^{n} H^4 e^{2\pi \nu}}{8\pi^3 n!} \frac{\Gamma(-n-i\nu)\Gamma(\frac{3}{2}+n+i\nu)}{\Gamma(2+n+i \nu))\Gamma(\frac{1}{2}-n-i\nu)} \notag \\[5pt]
&\hspace{90pt}\times\Gamma\left(\frac{3}{2}+n+2i\nu\right)\Gamma\left(\frac{3}{2}+n\right)\Gamma(-n-2i\nu) \ ,\\[5pt]
\tilde{c}_n(\nu)&=\frac{i H^4 (1+n)(1+2n)}{16\pi^4(n+1)!}\,\Gamma^2\left(\frac{1}{2}+n\right)\Gamma\left(\frac{3}{2}+n-i\nu\right)\Gamma\left(\frac{3}{2}+n+i\nu\right) \notag \\[5pt]
&\hspace{90pt}\times\Gamma(-n+i\nu) \Gamma(-n-i\nu) \ .
\end{align}
\end{subequations}

At first sight the series coefficients $\tilde{a}_n(\nu)$ and $\tilde{c}_n(\nu)$ do not look the same as the $a_n(\nu)$ and $c_n(\nu)$ from Mellin integrals in the previous section.  Nonetheless, one can show numerically that the two expressions are in fact identical. The analytical expressions for the series coefficients can also be matched by applying 
Euler's reflection formula
\beq
\Gamma(1-z) \Gamma(z)=\frac{\pi}{\sin \pi z}, \quad z \notin \mathbb{Z}
\eeq
repeatedly to some of the Gamma functions.

To summarize, we have shown that the contact terms obtained by using the  K\"all\'en-Lehmann representation is exactly the same as the one we had calculated from \cref{sec:mellin}. The anomalous dimension will then also match by following the derivation starting at \cref{eq:anom_step1}.

\section{Anomalous Dimensions From SdSET}
\label{sec:EFT}

Much of the confusion surrounding the physics of dS is that the long wavelength description in terms of scaling operators and scaling dimensions is obscured by the presence of Hankel functions. However, as we have seen, this results in an infinite series in $k\tau$, such that most terms are irrelevant in the long wavelength limit. Furthermore, the long distance behavior is not manifest at intermediate steps of the calculation, obscuring the physical origin of any non-trivial corrections. Calculations must typically be performed exactly as a function of $k$, which can be a major technical obstacle. This was evident in the previous sections where we had to provide the exact Mellin representation of the one-loop power spectrum, only to arrive at a result that look like an expansion in the long distance operators.

In this section, we will address the conceptual and technical complications by working with SdSET for principal series fields. The purpose of this effective theory is to make the nature of the superhorizon evolution manifest at the level of the effective fields, operators, and Lagrangian. The SdSET emerges as a power counting expansion in the limit $k/[a H] \ll 1$.
The EFT models the dynamics of IR modes with $k \ll aH$, so that the scale $aH$ can be interpreted as a time-dependent cutoff.
This implies that as the universe evolves in time, new modes enter the effective theory as they redshift to longer wavelength.

SdSET provides us with a language to clarify the puzzles that the computations of the anomalous dimensions introduce. We will see that new types of (non-dynamical) operators must be included in the IR description in order to reproduce the anomalous dimensions computed above.  These novel operators are additionally crucial to maintaining the positivity of the $\phi$ correlators, as we will discuss below.

\subsection{Principal Series SdSET}

The most straightforward way to derive the SdSET is to start with the full theory and perform a mode decomposition to identify the IR degrees of freedom.  We will start from the theory of a massive principal series scalar $\phi$, whose action is given by
\beq\label{eq:S_2phi}
S_{2,\phi} = \int \d t\, \d^3 x\, a^3(t) \frac{1}{2}\left(-\partial_\mu \phi\s \partial^\mu \phi - m^2 \phi \right) \ .
\eeq
We want to isolate the long wavelength behavior, so we can study the equations of motion for $\phi(\k =0, t)$ 
\beq
\ddot \phi(\k =0, t) + 3 H \dot \phi(\k =0, t) + m^2 \phi(\k =0, t) = 0 \ ,
\label{eq:keq0EOM}
\eeq
The resulting solutions are
\beq
\phi(\k = 0, t) = c_1 [aH]^{-\Delta} + c_2 [aH]^{-\bar \Delta} \ ,
\eeq
where $c_1$ and $c_2$ are constants that are determined by boundary conditions, and
\beq
\Delta = \frac{3}{2} + i \nu\ , \qquad \bar \Delta = \frac{3}{2} - i \nu\ , \qquad \nu = \sqrt{\frac{m^2}{H^2} - \frac{9}{4}} \ .
\eeq
This is, of course, the same behavior we found from taking the $k \to 0$ limit of the full solutions above. For our purposes, it is still useful to observe that the dimensions $\Delta$ and $\bar \Delta$ can be defined from the time-evolution of these two solutions, without ever determining the full $\k$-dependence.

Based on these solutions, we express the UV field $\phi$ as a mode expansion
\begin{equation}
\phi(\x,t)=H \big([aH]^{-\Delta}\vp(\x,t)+ [aH]^{-\bar \Delta }\vm(\x,t)\big)+\Phi_H(\x,t) \ ,
\label{eq:modeExpansionPhi}
\end{equation}
where $\vpm$ are the IR degrees of freedom and $\Phi_H$ are the UV modes with $k > aH$ that we will ultimately integrate out (which is trivial in a free theory). Since $\phi$ is a real field and $\Delta^* = \bar \Delta$, we must impose the constraint $\varphi_- = \varphi_+^*$. 

A useful aspect of this ansatz is that it implies that $\vpm$ transform as scaling operators under the dS isometeries, 
\begin{subequations}
\begin{align}
x^{i} &\s\to\s x^{i}-2\left(b_{j}\s x^{j}\right) x^{i}+b^{i}\left(\sum_{j}\big(\s x^{j}\s\big)^{2}-[a(\t)\s H]^{-2} \right) \ , \\ 
\t &\s\to\s \t+2\s b_{j}\s x^{j} \ .
\end{align}
\end{subequations}
Specifically, in order for $\phi$ to transform as a scalar, $\phi(\x,t) \to \phi(\vec{x}^{\s\prime},t')$, we must have
\begin{align}
\vpm(\x,\t) &\s\s\to\s\s \left[1-2\s\Delta_\pm\s x_i\s b^i + \big(x^{2}-[a\s H]^{-2}\big)\s b_i\s \partial^{i}-2\s x^{i}\s \vec{x} \cdot \vec{\partial} + 2\s b_i\s x^i\s \partial_{\t} \right]\s \vpm(\x,\t) \ ,
\end{align}
where $\Delta_+=\Delta$ and $\Delta_- = \bar \Delta$. In this regard, the fields $\vpm$ behave like operators in a CFT, which thus allows us to directly apply CFT intuition in the bulk of dS.

Substituting this ansatz back into \cref{eq:S_2phi}, we find the quadratic action for $\vpm$
\begin{align}
S_{2, \pm}=\frac{1}{2}\int \d^3 x\, \d \t \Big[ & {[a H]^{2 i\nu} \dot{\varphi}_{+}^2+[a H]^{-2 i\nu} \dot{\varphi}_{-}^2+2 \dot{\varphi}_{+} \dot{\varphi}_{-}-2 i\nu\left(\dot{\varphi}_{+} \varphi_{-}-\varphi_{+} \dot{\varphi}_{-}\right) } \\[3pt]
& -[a H]^{2 i\nu-2} \partial_i \varphi_{+} \partial^i \varphi_{+}-[a H]^{-2i \nu-2} \partial_i \varphi_{-} \partial^i \varphi_{-}-2[a H]^{-2} \partial_i \varphi_{+} \partial^i \varphi_{-}\Big] \ .\notag 
\end{align}
The dominant kinetic term in SdSET is first order in time, 
\beq
{\cal L}_{2,\pm} \supset - i\nu\left(\dot{\varphi}_{+} \varphi_{-}-\varphi_{+} \dot{\varphi}_{-}\right) \ ,
\eeq
whose equations of motion yield $\dvpm = 0$ up to power suppressed gradients. This correctly describes the expected $\vec{k} =0$ solutions as desired. As explained in~\cite{Weinberg:2008hq,Cohen:2020php}, the higher-order kinetic terms such as $\dvpm^2$ should be evaluated using the lower order equations of motion to avoid introducing additional degrees of freedom, and thus these terms are suppressed by powers of $(k/[aH])^2$.

Unlike EFTs commonly used in particle physics,\footnote{Many cosmological EFTs also require input of the initial conditions, including the EFT of LSS~\cite{Baumann:2010tm, Carrasco:2012cv}. It is also be required to match the time evolution beyond leading order in EFTs for classical systems~\cite{Beneke:2023ndu}.} the action alone does not determine the correlation functions of this effective theory. Instead, each mode enters the EFT at the time $t_\star$ when $k = a(t_\star) H$. The mode evolution prior to $t_\star$ is then encoded as stochastic initial conditions that are determined by matching to the full UV description.

We can determine the free theory matching, by starting with the mode expansion of the UV scalar field $\phi$ as given in \cref{sec:mellin} above:
\beq
\phi(\k,\tau)=\bar\phi_k(\tau) a_{\vec{k}}^{\dagger}+\bar\phi^*_k(\tau) a^{\vphantom{{\dagger}}}_{-\vec{k}} \ ,
\eeq
with
\begin{subequations}
\begin{align}
\bar\phi_k(\tau)&=\frac{\sqrt{\pi}}{2} e^{\frac{\pi \nu}{2}} H (-\tau)^{3/2} H_{i \nu}^{(2)}(-k \tau) \ , \\[3pt]   
\bar\phi^*_k(\tau)&=\frac{\sqrt{\pi}}{2} e^{-\frac{\pi \nu}{2}} H (-\tau)^{3/2} H_{i \nu}^{(1)}(-k \tau) \ .
\end{align}
\end{subequations}
Repeating the series expansion around $-k \tau \to 0$ yields the leading terms
\beq\label{eq:series_rep}
\bar\phi_k(\tau) =\frac{1}{k^{3/2}} \left(\frac{i 2^{i \nu } e^{\frac{\pi  \nu }{2}} \Gamma (i \nu ) }{2\sqrt{\pi }} H(-k\tau)^{\bar \Delta}+\frac{i 2^{-i \nu } e^{-\frac{\pi  \nu }{2}} \Gamma (-i \nu )}{2\sqrt{\pi }} H (-k\tau)^{\Delta} \right) \ ,
\eeq
so that we can identify
\begin{subequations}
\bea
\vp(\k) &=& \frac{k^{\Delta}}{\sqrt{2} k^{3/2}}\left( C_\nu a_{\vec{k}}^{\dagger}+ D_\nu^* a_{-\vec{k}} \right) \ ,\\[3pt]
\vm(\k) &=& \frac{k^{\bar \Delta}}{\sqrt{2} k^{3/2}} \left( D_\nu a_{\vec{k}}^{\dagger}+ C_\nu^* a_{-\vec{k}} \right) \ ,
\eea
\end{subequations}
where
\beq
C_\nu =\frac{i 2^{-i \nu } e^{-\frac{\pi  \nu }{2}} \Gamma (-i \nu )}{\sqrt{2\pi }}\ , 
\qquad \text{and}\qquad
D_\nu = \frac{i 2^{i \nu } e^{\frac{\pi  \nu }{2}}  \Gamma (i \nu )}{\sqrt{2\pi }}  \ .
\eeq
The power-law behavior in $k$ for $\vp$ and $\vm$ is consistent with both being scaling operators of dimension $\Delta$ and $\bar \Delta$ under the dS isometeries. 

Given this definition terms of the UV creation and annihilation operators, it is easy to check that $[\vp,\vp]=[\vm,\vm]=0$ and
\beq
[\vp(\k\s),\vm(\kp)] = -\frac{1}{2\nu} (2 \pi)^3\delta(\k+\kp) \ .
\eeq
Of course, these results would also follow from the EFT desciption alone, as the canonical commutation relation $[\vp,\Pi_+] = i\s (2 \pi)^3\delta(\k+\kp)$, where $\Pi_+ = -2i \nu \vm$ is the conjugate momentum to $\vp$. IN, the two point functions of $\vpm$ are non-zero and determined by the UV theory, reflecting the fact that their initial conditions are classical statistical random variables. Matching the equal time two point function of the free theory requires, for example, that
\beq
\big\langle \vp(\k\s) \vp(\kp) \big\rangle = \frac{k^{2\Delta}}{2 k^3} \frac{1}{2\nu \sinh(\pi\nu)}(2 \pi)^3\delta(\k+\kp) \ .
\eeq
In the large mass limit, $\nu \to m/H$ and $\langle \vp(\k\s) \vp(\kp) \rangle \propto e^{-\pi m/H}$, consistent with a Boltzmann suppression of the fluctuations of heavy fields in dS at a temperature\footnote{The missing factor of two in the exponential relative to the Boltzmann suppression can be understood from the fact that the density is the absolute value squared of the wavefunction.} $T = H/(2\pi)$.

To summarize, we have shown that $\vp$ and $\vm$ in the free theory behave as scaling operators with dimensions $\Delta$ and $\bar \Delta$ respectively. The non-zero commutator between $\vp$ and $\vm$ is consistent with scale invariance because $\Delta + \bar \Delta =3$. Since $\bar \Delta = \Delta^*$, this requires\footnote{The condition $\Delta +\bar \Delta =3$ and $\bar{\Delta} =\Delta^*$ can also be derived from the bottom up following precisely the same procedure described in~\cite{Cohen:2020php}. The two are related, as $\Pi_+ \propto \vm$ is a consequence of the leading kinetic term and thus the dimensions are fixed by the canonical commutator and/or the scaling of the kinetic term. Either way, one finds the main result that $\gamma =0$ in absence of interactions within SdSET.} that ${\rm Re}\s\Delta = 3/2$. This statement is directly related to the fact that the unitary states associate with the principal series operators are labeled by eigenvalues under dilatation $\Delta$, with ${\rm Re}\, \Delta = 3/2$.

\subsection{Interactions in Principal Series SdSET}

One of the central motivations for SdSET is to make behavior under time-evolution manifest from dimensional analysis. Intuitively, we are working in an EFT where our UV scale is $\Lambda(\t) = a(\t) H$.  It is therefore natural to expect the size of any interaction to be controlled by dimensional analysis.
This in turn determines the scaling of corrections in terms of $k/\Lambda(\t)$, and thus ties the power counting to the time evolution.

Our EFT expectation can be shown to follow directly from basic physics principles. For one, any physical infinitesimal distance depends on $a(\t) \times \d\x$.  This implies that there is a trivial rescaling symmetry $a(\t)\to \lambda a(\t)$ and $\x \to \lambda^{-1} \x$ ($\k \to \lambda \k$) that leaves physical quantities unchanged. Under this symmetry, the free fields $\vp$ transform as scaling operators with dimensions $\Delta$ and $\bar \Delta$. As a result, we can determine powers of $a(\t)$ associated with any interaction just from dimensional analysis in the action. By construction, the units and scaling dimensions of $\vpm$ are the same.  Corrections therefore always arise in the combination $a(t) H$ to maintain that the action is dimensionless.

Given these points of view, we can construct interactions according to the following power counting rules: we assign $\vp$ and $\vm$ dimensions $\Delta$ and $\bar \Delta$, $\d^3x$ has dimension $-3$ and $\d\t$ has dimension 0. Powers of $\Lambda(\t) = a(\t) H$, which has dimension 1, are then introduced to make the action dimensionless. If the power $\Lambda(\t)$ that appears in an interaction is positive (negative) we call the interaction relevant (irrelevant) as it become more (less) important as time is evolved forward. Dimension three interactions  have no powers of $\Lambda(\t)$, and are therefore called marginal.

In the free theory, the principal series fields that have dimension $\Delta, \bar \Delta = \frac{3}{2}\pm i\nu$. Therefore, when thinking about interactions in the interaction picture, we can use these values for power counting. However, as imaginary dimensions do not produce growing or decaying time evolution, we will define relevant, marginal, and irrelevant solely in terms of the real powers of $\Lambda(\t)$, which are determined by the real part of the dimensions of operators. Applying this power counting, the leading polynomial interactions take the form 
\beq
S_{\text {int }} \supset \int \d^3 x\, \d \t[a H]^{3-n \Delta-m\bar \Delta} \frac{c_{n, m}}{n !m!} \varphi_{+}^n\varphi_-^m  \ ,
\eeq
where $c_{n,m} = {\cal O}(1)$ are dimensionless coefficients that can be determined from matching. For $n+m \geq 3$, $(3-n\Delta -m \bar \Delta) = \frac{3}{2}(2-n-m)+(m-n)i\nu$, hence the real part of $3-n\Delta -m \bar \Delta$ is always strictly less than zero.  We conclude that all polynomial interactions are irrelevant.\footnote{Recall that one relies on field redefinitions to eliminate spurious relevant interactions in the the case of complementary series fields~\cite{Cohen:2020php} when deriving the SdSET interactions.  In the case of principal series fields, the interactions are automatically irrelevant so one does not need to perform any additional field redefinitions.}

This conclusion would seem to be at odds with our results from the previous sections. In SdSET, when $\Delta = \bar \Delta^*$ is complex, the quadratic action requires that with ${\rm Re}\s \Delta = {\rm Re}\s \bar \Delta =\frac{3}{2}$. Therefore, any real anomalous dimensions would have to come from marginal or relevant interactions in SdSET. The absence of any such interaction is inconsistent with the real anomalous dimensions for $\vp$ and $\vm$ that we found in the previous section. We cannot resolve this issue by simply declaring that the corrections to $\Delta$ are determined in the UV by matching, since this would be inconsistent with the canonical commutations relations that can be derived within SdSET in the IR.

The resolution to this apparent contradiction is that our EFT is missing additional operators. Specifically, suppose our theory contains an operator ${\cal O}(\t,\x)$ with dimension $\Delta_\O$ such that we can introduce additional interactions
\beq
S_{\text {int }} \supset \int \mathrm{d}^3 x \operatorname{dt}[a H]^{\frac{3}{2} - \Delta_\O}\left( c_{\cal O} [aH]^{-i \nu} \vp+ c^*_{\cal O} [aH]^{i \nu} \vm \right) {\cal O}(\t,\x)\ ,
\eeq
If ${\rm Re} \Delta_\O \leq \frac{3}{2}$, this operator can potentially generate corrections and do not vanish at late times. While this may explain the anomalous scaling in the coupling of heavy fields to light fields, it seems for self-interacting heavy fields we are missing to operators ${\cal O}$ with the right scaling behavior. However, one hint is that if we have a second principal series field $\sigma$, then ${\cal O} =\sigma_\pm$ 
could
have the right dimensions to be marginal.  We will explain how this observation resolves the puzzle stated above, and allows us to reproduce the calculation of the anomalous dimensions for $\phi$ within SdSET.

\subsection{UV Composite Operators and IR Contact Terms}

One can simply define composite operators within SdSET as the product of the local operators $\vpm$. In the free theory, an operator of the form
\beq
{\cal O}_{p,q}(\x) =[\vp^p \vm^q](\x) \qquad\text{with}\qquad {\cal O}_{p,q}(\k) = \int \d^3 x\, e^{i \vec{k} \cdot \x}{\cal O}_{p,q}(\x) \ ,
\eeq
will have a dimension
\beq
\Delta_{p,q} =  p \Delta + q\bar \Delta \ .
\eeq
This property is maintained at loop level, provide we use a scaleless regulator.

Above, we derived \cref{eq:single_sum}, which shows that the two-point function of $\sigma^2$, where $\sigma$ is a  UV principal series field, contains a series of non-trivial contact (${\cal O}(k^0)$) terms, 
\begin{align}
\langle  \sigma^2(\k,\tau) \sigma^2(\kp,\tau') \rangle' \supset   &\sum_n a_n(\nu) (-\tau)^{2 i\nu+2n+3}(-\tau')^{-2i\nu -2n} +(\nu \rightarrow -\nu)\nonumber\\[3pt] 
&+\sum_n c_n(\nu) (-\tau)^{2n+3}(-\tau')^{ -2n} \ .
\end{align}
For this to be consistent with the scaling with dS isometeries, we need to identify new operators $\O_{j,n}$ and $\bar \O_{j,n}$ with dimensions
\begin{align}
\Delta_{j,n} = 3 + 2i  j \nu_\sigma + 2n\ , 
\qquad\text{and} \qquad
\bar \Delta_{j,n} = - 2i j \nu_\sigma - 2n = 3- \Delta_{j,n} \ ,
\end{align}
where $j =0,\pm 1$ and $n =0,1,2,..$ (precisely as in \cref{eq:comp_2pt_jn}) such that
\begin{align}\label{eq:s2_EFT}
\sigma^2(\x)  \supset \sum_{j,n} H^2 \Big([aH]^{-\Delta_{j,n}} &\O_{j,n}(\x) + [aH]^{-\bar \Delta_{j,n}}\bar \O_{j,n}(\x) \Big)\ .
\end{align}
Based on the scaling dimensions, the operators $\O_{j,n}$ could be written in terms of $\vpm$ and derivatives within SdSET. For example, matching the dimensions would give us $\O_{j=\pm,n=0} = \sigma_\pm^2$ so that as $\Delta_{+,n=0} = 2 \Delta$ and $\Delta_{-,n=0} = 2\bar \Delta$. However, the shadow operators $\bar \O_n$ cannot be constructed from any composite operator in SdSET. Yet, we see that they must be present. In order to match the commutator in \cref{eq:sig2_com}, we must have
\beq
\langle [\O_{j,n}(\x,\t), \bar \O_{j,n}(0,\t')] \rangle =
\theta(\t-\t')\delta(\x)H^{-4} \Big( C(\Delta_{j,n}) - C^*(\Delta_{-j,n})\Big)\ .
\eeq
Yet, at separated points, we see no contributions to the two-point functions that have dimension $2 \bar \Delta_{j,n}$ and therefore we also require that 
\beq
\langle \bar \O_{j,n}(\k\s) \bar \O_{j,n}(\kp) \rangle = 0 \ .
\eeq
Specifically, we can check that there are no terms that are non-analytic in $k$ with this scaling behavior from the UV loop calculation. Therefore, in order to reproduce the ``contact terms" of the UV theory, we must introduce shadow operators $\bar \O$ that have no long distance correlations. 

Now that we have introduced these additional operators, we can consider the anomalous dimension calculation in the language of SdSET. In the UV theory, we have two field $\phi$ and $\sigma$, which in SdSET become $\vpm$ and $\sigma_\pm$, with dimensions $\dimPhi$, $\dimBPhi$,$\dimSig$, and $\dimBSig$ respectively. The UV theory contains the interaction $\H_{\rm int} = \lambda H \phi \sigma^2$, which in order to match must include
\beq\label{eq:comp_hint}
{\cal H}_{\rm int} \supset \lambda H^4 ([aH]^{-\dimPhi} \vp + [aH]^{-\dimBPhi} \vm)\sum_{j,n}([aH]^{-\Delta_{j,n} }\O_{j,n} + [aH]^{-\bar \Delta_{j,n}} \bar \O_{j,n}) \, .
\eeq
Repeating our perturbative calculation of the $\phi \to \vp$ power spectrum, we have
\begin{align}
\langle \vp(\k\s) \vp(\kp) \rangle &\supset  -\lambda^2\ \langle \vp(\k\s) \vp(\kp) \rangle_{\lambda=0} \sum_{j,n}\int \d \t_2\int^{\t_2} \d \t_1 \frac{[aH]_2^{3-\bar{\Delta}_\phi}}{2\nu_{\phi}} \notag\\[2pt]
&\hspace{60pt}\times\Big([aH]_2^{-\Delta_{j,n}}[aH]_1^{3-\Delta_\phi-\bar{\Delta}_{j,n}} [\bar \O_{j,n}(\t_1), \O_{j,n}(\t_2)] \notag\\[2pt]
&\hspace{85pt}+ [aH]_2^{-\bar {\Delta}_{j,n}}[aH]_1^{3-\Delta_\phi-{\Delta}_{j,n}} [ \O_{j,n}(\t_1), \bar \O_{j,n}(\t_2)]\Big)\ ,
\end{align}
where the operators appearing inside the commutators are in momentum space, and we only include time dependence here for brevity since these commutators are momentum independent.  Note that because of the commutator only includes a single time ordering, only the first commutator contributes to the integral since $\t_1 < \t_2$. As a result, we find
\begin{align}
\langle \vp(\k\s) \vp(\kp) \rangle \supset &  -\frac{\lambda^2}{2\nu_{\phi}H^4}\langle \vp(\k\s) \vp(\kp) \rangle_{\lambda=0} \notag \\
&\times\int \d\t_2 [aH]_2^{3-\Delta_\phi-\bar{\Delta}_\phi}\left(\sum_{j,n}\frac{C(\Delta_{j,n})^*}{3-\Delta_\phi-\bar{\Delta}_{-j,n}}-\frac{C(\Delta_{j,n})}{3-\Delta_\phi-\bar{\Delta}_{j,n}}\right)\ ,
\end{align}
From here we can see that the $\t_2$ integral is diverges, as $\dimPhi+\dimBPhi =3$ and gives us precisely the same logathrimic term as in~\cref{eq:anom_step1}, along with the anomalous dimension of $\vp$ and, similarly, $\vm$.

With the introduction of the operators $\bar \O_n$ into SdSET, the origin of the anomalous dimension again follows from power counting. Crucially, since ${\rm Re} \Delta > 3/2$ and ${\rm Re}\bar \Delta < 3/2$, the interaction Hamiltonian \cref{eq:comp_hint} contains a relevant interaction. This is in contrast to the composite operators formed from the dynamical SdSET fields alone, whose interactions are always irrelevant. Moreover, the expansion in terms of the SdSET fields, including $\vpm$, automatically isolates the terms of interest in the calculation without the need to keep track of all the terms in the expansion of the mode functions of $\phi$.

We will now see that these novel shadow operators are also critical to reproducing the expected positivity properties of the $\phi$ correlators.  This provides even more evidence that these operators must be included in the SdSET description for consistency.

\subsection{Physical Origin of Contact Operators}

Naturally, one would also like to understand why SdSET needs additional operators that are not manifest from the field content. At face value, it might seem like a break-down of the EFT description that there are contributions to the correlators that are not visible from the long-wavelength field content.

Essentially, the issue comes down to an order of operations.  The question is if one can expand the integrand before integrating (essentially performing a method of regions decomposition~\cite{Beneke:1997zp, Beneke:2023wmt}), or is it required to perform the full integral first.  
Take the loop integral over the principal series field in \cref{eq:mellin_loop}. It we were to evaluate the expression on the poles corresponding to composite operators in SdSET, $2s_1 +2s_2 \to 3-\Delta_n = (0,\pm) 2 i\nu + 2 n$ and $2s_3 + 2 s_4 = 3-\Delta_m = (0,\pm) 2 i\nu - 2 m$, then we would find
\beq
\int \frac{\d^3 p}{(2 \pi)^3}(-\tfrac{i}{2} p \tau)^{2 \Delta_n-3} (\tfrac{i}{2}p \tau')^{2\Delta_m-3} = 0 \ .
\eeq
This is a reflection of the fact that in the series expansion, any $k^0$ contribution to the $\sigma_\pm^2$ power spectrum would contribute a power law divergence (by dimensional analysis) and thus would vanish when using a scaleless regulator.  In other words, some additional UV input is required to fully determine the anomalous dimensions in the EFT, which is encoded in the coefficient of the non-dynamical operator introduced in the previous section.

In prior work~\cite{Marolf:2010zp,DiPietro:2021sjt,Hogervorst:2021uvp,Bros:2010rku}, the appearance of the real anomalous dimension for $\phi$ was given the interpretation of the width due to the decay\footnote{The same point of view is also useful in both dS and AdS for recovering the flat space $S$-matrix from the correlators in curved space.} of $\phi \to 2\sigma$. Since energy is not conserved in dS, this process is not forbidden by kinematics. Since $\gamma \neq 0$ implies that no operator has a dimension corresponding to a unitary state, it has been suggested that the presence of the anomalous dimension is consistent with idea that $\phi$ is merely a resonance, like a decaying particle in flat space, and thus does not appear in the asymptotic Hilbert space.

However, from a purely long wavelength perspective, this interpretation remains somewhat unsatisfying. In flat space, the resonance language is useful to distinguish fields that do and do not appear in the S-matrix as asymptotic states. In dS, we can generate these anomalous dimension in the case of a signal field $\phi$ with a cubic self-interaction $g \phi^3$. Therefore, we will generally expect interacting theories in dS contain no fields that correspond to states and thus everything is a ``resonance" in this sense. Instead, from the long wavelength perspective, the benefit of SdSET is that it allows one to work directly with operators, without needing to interpret the space of asymptotic states. 

From the operator perspective, the remaining question is why we should include operators that are not described in terms of the long wavelength (EFT) degrees of freedom. The answer is that it is a consequence of prioritizing symmetry and power-counting over minimal operator content. Specifically, we insistent that contact terms, in the sense of terms that are analytic in $\vec{k}$, must be consistent with the de Sitter invariance associate with the scaling dimensions of the corresponding operator. Contact terms for generic operators in QFT are often incalculable and therefore one could simply introduce these terms by matching.  

We could therefore include terms like 
\beq
\langle \sigma_+^2(\k,\t) \sigma_+^2(\kp,\t')\rangle'=  c_2\s k^{4 \dimSig -3} + d_2\s k^0 [a(\t') H]^{3 - 4\dimSig} +\ldots \ .
\eeq
The second term is naively forbidden by the conformal Ward identities if we hold the dimension of $\sigma_+^2$ fixed.  It is well known that quantum field theories can have anomalies in the form of contact terms that break the classical symmetries. In this case, thinking of such terms as anomalous is unhelpful as the UV operators do not have fixed dimensions and therefore these terms do not actually break the symmetry. Instead, a better interpretation is that these operators correspond to dS invariant contact terms that cannot be associated with the dimension of an operator whose scaling dimensions are measurable at separated points.

This strategy of introducing new non-dynamical modes to incorporate UV contact terms has precedent in other modern EFTs~\cite{Manohar:1993qn,Liu:2008cc,Bauer:2010cc,Rothstein:2016bsq}. For example, in Soft Collinear Effective Theory (SCET), there are factorization violating observables for which one must introduce so-called Glauber modes separately in order to correctly match the dynamics of the UV theory~\cite{Liu:2008cc,Bauer:2010cc,Rothstein:2016bsq}. These modes are localized in time but cannot be integrated out. Moreover, these modes have a unique power counting compared to the other modes in SCET.  Thus the EFT power counting rules demand that these modes are treated separately. This is very much the same situation we find in SdSET.  We therefore conclude that these new SdSET contact terms correspond to distinct operators that are required in order to make power counting explicit.  We leave the systematic classification of such operators using \emph{e.g.}~the method of regions in dS~\cite{Beneke:2023wmt} for future work.

\subsection{Revisiting Positivity in dS}
A real scalar field $\phi(\x,t)$ should obey a number of (apparently) trivial positivity constraints. The most obvious such constraint is that the power spectrum should be positive. Concretely, if we calculate the wavefunction of the universe in the $\phi$ basis, $\Psi[\phi(\k\s),t]$, then the power spectrum is given by
\beq
\langle \phi(\k,t) \phi(\kp,t) \rangle' = \int {\cal D}\phi |\phi(\k)|^2 |\Psi[\phi(\k),t]|^2 >0  \ ,
\eeq
where we used the reality condition $\phi(\k\s)^* = \phi(-\k\s)$.

This positivity condition is clearly satisfied for free fields. In the principal series, recall that
\begin{align}
\langle \phi(\k,t) \phi(\kp,t) \rangle' &\simeq  \frac{H^2 (-\tau)^3}{2\pi}\bigg( \Gamma(-i \nu)^2 \cosh(\tfrac{\pi \nu}{2} )\left(-\tfrac{k \tau}{2} \right)^{i 2 \nu} + \Gamma(i \nu)^2 \cosh(\tfrac{\pi \nu}{2} ) \left(-\tfrac{ k \tau}{2}\right)^{-i 2\nu}  \notag\\[3pt]
&\hspace{85pt}+\Gamma(-i \nu) \Gamma(i \nu) \cosh(\pi \nu) \bigg) \ .
\end{align}
Although the first two terms are oscillatory in $k$, the third term is manifestly positive and it is also larger that the other two terms. As a result, this expression is positive as expected. This is also what we would find in SdSET if we use 
\beq\label{eq:phi_EFT0}
\phi(\x,t) \to H \big([aH]^{-\Delta}\vp(\x,t)+ [aH]^{-\bar \Delta }\vm(\x,t)\big) \ .
\eeq
Since $\vp^\dagger = \vm$ in SdSET, the expression on the right-hand side is also real and therefore should be positive, as it must be from matching onto the free theory in the UV.

Now suppose that $\vp$ and $\vm$ acquire anomalous dimensions so that $\dimPhi = \frac{3}{2} + i \nu + \gamma$ and $\dimBPhi = \dimPhi^*$ with ${\rm Re} \gamma \neq 0$. Since the dimensions are related by complex conjugation, we would naively expect the combination in \cref{eq:phi_EFT0} to correspond to the real field $\phi$ as well, as it did in the free theory.  However, if we define $\phi$ by \cref{eq:phi_EFT0}, then by dS symmetry, the power spectrum at leading order in $k/[aH]$ is given by
\beq
\langle \phi(\k,t) \phi(\kp,t) \rangle' \stackrel{?}{\simeq} [a(t)H]^{-3} \left( c \left(\frac{k}{aH}\right)^{2\gamma+2 i\nu} +c^* \left(\frac{k}{aH}\right)^{2\gamma^*-2 i\nu}\right) \ ,
 \eeq
where $c$ is some complex valued constant determined by 1-loop matching. Note that there is no longer a $k^0$ contact term as this is only allowed by the conformal Ward identities when $\dimPhi + \dimBPhi = d=3$.  Without the $k^0$ term this expression is cannot be made manifestly positive in the $k \to 0$ limit, which is inconsistent with the fact that $\phi$ is a real field.

The resolution to this paradox is that $\phi$ is not a scaling operator. As a result, the contact terms that appear in the two-point function of the UV field $\phi$ are not fixed by the lowest dimension scaling operators that contribute at {\it separated points}, namely $\vp$ and $\vm$.  These SdSET fields are not real, they are thus the individual power spectra of SdSET are not required to be positive. Moreover, once we include the couplings to other operators, it is not even clear that $\vp$ and $\vm$ need to be related by complex conjugation. Specifically, the anomalous dimension crucially requires that $\vpm$ interact with $\O_{j,n}$ and $\bar \O_{j,n}$. The real field $\phi$ is therefore a linear combination of all the scaling operators $\vpm$,  $\O_n$, and $\bar \O_n$, namely
\beq
\phi(\k,t) \sim H [aH]^{-3}\left(b_+ \vp [aH]^{-\dimPhi} + b_-\vp [aH]^{-\dimBPhi}  + \sum_{j,n}\Big( b_n \O_{j,n} + \bar b_{n} \bar \O_{j,n}\Big)\right) \ .
\eeq
In fact, from \cref{eq:s2_com} we can see that 
\begin{align}
\langle \phi(\k,t) \phi(\kp,t) \rangle' \supset& -\lambda^2 H^4\int_{-\infty}^{\tau}\frac{\d\tau_2}{(-H \tau_2)^4} \big[\phi(\k,\tau_2),\phi(\kp,\tau)\big] \notag\\[4pt]
&\hspace{12pt}\times\int_{-\infty}^{\tau_2} \frac{\d \tau_1 }{(-H \tau_2)^4} \big[\phi(\kp,\tau_1),\phi(\k,\tau)\big]\langle\sigma^2(\k,\tau_1)\sigma^2(\kp,\tau_2)\rangle \ .
\end{align}
Note that our previous results have shows that the right-hand side includes a $k^0$ term, due the fact that both the commutators and $\langle [\sigma^2](\k) [\sigma^2](\kp) \rangle$ contain non-zero contact terms. Therefore, the matching requires that we include an extra term $d\times k^0$ such that 
\beq
\langle \phi(\k,t) \phi(\kp,t) \rangle' \to  [a(t)H]^{-3} \left( d + c \left(\frac{k}{aH}\right)^{2\gamma+2 i\nu} +c^* \left(\frac{k}{aH}\right)^{2\gamma^*-2 i\nu}\right) \ .
\eeq
where $d \propto 2\sum_{j,n} \langle \O_{j,n}\bar \O_{j,n}\rangle$. Under the conditions that ${\rm Re}  \, \gamma > 0$, dominant contribution (in the limit $k\ll aH$) to this two point function is the contact term, $d$, and the two point function remains positive as long as $d>0$. The case ${\rm Re} \, \gamma <0$ and/or $d \leq 0$ cannot be resolved in the same way and are forbidden by these kinds of classical positivity arguments~\cite{Green:2023ids}.

The behavior is encoded directly in the K\"all\'en-Lehmann representation. Including the anomalous dimension, the fields $\phi$ can be still be decomposed in terms of a spectral density
\beq
\langle \phi(\k,t) \phi(\kp,t) \rangle' =  \int_{\mathbb{D}} \frac{\d \nu}{2 \pi} \rho_\phi\left(\frac{d}{2}+i \nu\right) G(\k ; \nu) \ .
\eeq
Since $G(\k; \nu)$ is an equal-time two point function of a principal series field, $G(\k;\nu) > 0$ and the spectral density $\rho_\phi\left(\frac{d}{2}+i \nu\right) \geq 0$ for all $\nu$~\cite{Hogervorst:2021uvp,DiPietro:2021sjt}.

\section{Conclusions}\label{sec:con}

Understanding physics in de Sitter space has long been limited by a mix of conceptual and technical obstacles~\cite{Flauger:2022hie}. The natural observables associated with the long wavelength behavior of fields in accelerating cosmologies are very different from the local scattering experiments that are familiar from flat space quantum field theory~\cite{Witten:2001kn,Bousso:2004tv}.
In addition, perturbative calculations are technically challenging~\cite{Akhmedov:2019cfd,Green:2022ovz}, which makes the need for physical principles that can be used to guide intuition essential~\cite{Baumann:2022jpr}.

In this paper, we provided a new perspective on the origin of anomalous dimensions for principal series field. We showed how they arise by direct calculation using the Mellin representation, which confirms the results previously calculated using the K\"all\'en-Lehmann representation~\cite{Bros:2010rku,Marolf:2010zp}. 
We then demonstrated how to reproduce the same result in SdSET. We showed how the appearance of the anomalous dimension relies on the presence of non-dynamical operators with unique scaling behavior. These terms are essential to the calculation of the scaling dimensions, and also to ensure that the power spectra of real fields are positive.

In the process, we have further developed the use of Mellin space as a tool to understand cosmological correlators. While there has been significant progress pure dS calculations using the K\"all\'en-Lehmann representation~\cite{Loparco:2023rug,DiPietro:2023inn,Loparco:2023akg}, inflationary backgrounds break the de Sitter isometeries~\cite{Cheung:2007st,Creminelli:2012ed,Hinterbichler:2012nm} and make defining a complete basis of states more challenging. The Mellin representation has no such limitation~\cite{Sleight:2019hfp,Qin:2022fbv,Qin:2022lva,Qin:2023bjk}, so the techniques developed here are directly applicable to inflationary models as well. Moreover, the Mellin space calculations  have a natural interpretation in the language of the SdSET, which therefore exposes many features that help simplify the understanding of loop corrections.

There remains much to understand about operator mixing both in the context of dS and inflation. Interacting light fields display much more complex behavior. In the massless limit, an infinite number of operators can mix~\cite{Baumgart:2019clc,Green:2020txs}, giving rise to stochastic inflation.  Determining non-trivial corrections to stochastic inflation is then equivalent to computing anomalous dimensions that correct this operator mixing~\cite{Cohen:2021fzf,Cohen:2021jbo,Cohen:2022clv} (see also~\cite{Gorbenko:2019rza,Mirbabayi:2020vyt}). These light fields can also mix with heavy fields to produce novel cosmological collider signals that have only recently begun to be explored~\cite{Lu:2021wxu,Chakraborty:2023qbp,Chakraborty:2023eoq}. Furthermore, at a practical level, a systematic scheme for regulating and renormalizing the calculation of cosmological correlators remains a work in progress~\cite{Premkumar:2021mlz,Beneke:2023wmt}.  It would be illuminating to show that the non-dynamical SdSET operators introduced here could be identified using a method-of-regions style analysis~\cite{Beneke:2023wmt}. In addition, we would like a better non-perturbative understanding of the constraints on cosmological correlators, including an argument that all divergences can be renormalized by counterterms for the scaling operators. Much like the study of scattering amplitudes~\cite{Bern:2022jnl}, progress in direct calculations informs the bootstrap approach, and vice versa. Therefore, the pursuit of a deeper understanding will likely require a combination of new computational tools and new conceptual insights.

\paragraph{Acknowledgements}
We are grateful to Priyesh Chakraborty, Victor Gorbenko, Kshitij Gupta, Austin Joyce, Manuel Loparco, Aneesh Manohar, Akhil Premkumar, Chia-Hsien Shen, John Stout, and Guanhao Sun for helpful discussions. TC is supported by the US~Department of Energy under grant \mbox{DE-SC0011640}.  DG and YH are supported by the US~Department of Energy under grant~\mbox{DE-SC0009919}.


\newpage
\appendix

\section*{Appendix}
 
\section{Series Sum}\label{app:sum}
In this appendix, we fill in some details for the calculation in \cref{sec:oneLoopFromMellin}, by explaining how to manipulate the triple series sum in \cref{eq:triple_sum} into a form that can be directly used to compare our results with the  K\"all\'en-Lehmann spectral representation. For brevity, we will only consider the case $\tau > \tau'$:
\begin{align}
\langle  \sigma^2(\k,\tau) \sigma^2(\kp,\tau') \rangle' \to&  - \frac{H^4}{32\pi^4 } \sum_{\pm_1,\pm_2,\pm_3 } \sum_{n_1,n_2,n_3=0}^\infty \Gamma\left(s_4+\frac{i \nu}{2}\right) \Gamma\left(s_4-\frac{i \nu}{2}\right)\notag\\[4pt]
&\times\frac{(-1)^{n_1+n_2+n_3} }{n_1!n_2! n_3!} \Gamma(\mp_1i \nu -n_1) \Gamma(\mp_2i \nu -n_2)\Gamma(\mp_3i \nu -n_3) \\[4pt]
&\times\left(-\tfrac{i}{2} \tau\right)^{3+(\pm_1 1\pm_2 1)i \nu +2(n_1+n_2)}\left(\tfrac{i}{2} \tau'\right)^{-(\pm_1 1\pm_2 1)i \nu -2(n_1+n_2)} \notag \ ,
\end{align}
where
\beq
s_4 = \frac{3}{2}+ i (\pm_1 1\pm_2 1\pm_31) \frac{\nu}{2} +n_1+n_2+n_3 \ .
\eeq
As an example, we are going to set $\pm_1\to +$, $\pm_2\to -$ and sum over $\pm_3$. This will give rise to part of the term $\sum_n c(n) (-\tau)^{2n+3}(\tau')^{-2n}$, the other two terms in the final sum can also be obtained using the same manipulations.

First notice that the exponents of $\tau, \tau^\prime$ only depend on the sum of $n_1+n_2$, therefore it is possible to rearrange the sum into a form that does not involve explicitly summing over powers of $\tau$ and $\tau^\prime$. If we set $n= n_1+n_2$ and change the summation bounds accordingly, we get:
\begin{align}
&\sum_{n=0}^{\infty}(-1)^{n}\left(-\tfrac{i}{2} \tau\right)^{ 2n}\left(\tfrac{i}{2} \tau'\right)^{ -2n- 3} \left(\sum_{n_1=0}^{n_1=n}\frac{1 }{n_1!(n-n_1)!} \Gamma(-i \nu -n_1) \Gamma(i \nu -n+n_1) \right)\notag\\[5pt]
&\hspace{20pt}\times \left(\sum_{\pm_3}\sum_{n_3=0}^\infty \frac{(-1)^{n_3}}{n_3!}\Gamma(\mp_3 i \nu -n_3) \Gamma\left(\frac{3}{2}+n+n_3\pm_3 i\nu\right) \Gamma\left(\frac{3}{2}+n+n_3\right)\right)\ .
\end{align}
For fixed $n$, the sum over $n_3$ and $n_1$ factorizes. Both sum can be performed using the series representation of the hypergeometric function $_2F_1(a,b,c,z)$. For the sum over $n_3$ we find
\begin{align}
&\sum_{\pm_3}\sum_{n_3=0}^\infty \frac{(-1)^{n_3}}{n_3!}\Gamma(\mp_3 i \nu -n_3) \Gamma\left(\frac{3}{2}+n+n_3\pm_3 i\nu\right) \Gamma\left(\frac{3}{2}+n+n_3\right)\notag\\
&=\frac{2^{-1-2n}\pi^{5/2}\Gamma(\frac{3}{2}+n)}{\left(1+\cosh(2\pi\nu)\right)\Gamma(n+2)\Gamma(-\frac{1}{2}-n-i \nu)\Gamma(-\frac{1}{2}-n+i \nu)}
\end{align}
The finite sum over $n_1$ is 
\begin{align}
\sum_{n_1=0}^{n_1=n}\frac{\Gamma(-i \nu -n_1) \Gamma(i \nu -n+n_1)}{n_1!(n-n_1)!}   
= \frac{(-1)^n 2^{2n}\pi^{3/2}\text{csch}(\pi \nu)^2\Gamma(\frac{1}{2}+n)}{n! \Gamma(n+1-i\nu)\Gamma(n+1+i\nu)} \ .
\end{align}
This is used in deriving Eqs.~(\ref{eq:coeffsSinglesum}) above.

\phantomsection
\addcontentsline{toc}{section}{References}
\small
\bibliographystyle{utphys}
\bibliography{PS_EFT}

\end{document}